\documentclass[numberedappendix]{emulateapj}

\slugcomment{Accepted Version}

\shorttitle{Age Determination of Six Intermediate-age SMC Star Clusters with HST/ACS}
\shortauthors{Glatt et al.}

\begin{document}

\title{Age Determination of Six Intermediate-age SMC Star Clusters with HST/ACS
\altaffilmark{*}}

\author{Katharina Glatt\altaffilmark{1,2,3}, Eva K. Grebel\altaffilmark{3},
Elena Sabbi\altaffilmark{3,4}, John S. Gallagher III.\altaffilmark{2}, 
Antonella Nota\altaffilmark{4},  
Marco Sirianni\altaffilmark{4}, Gisella Clementini\altaffilmark{5}, 
Monica Tosi\altaffilmark{5}, Daniel Harbeck\altaffilmark{2},
Andreas Koch\altaffilmark{6}, Andrea Kayser\altaffilmark{1},
and Gary Da Costa\altaffilmark{7}}
\altaffiltext{*}{Based on observations made with the NASA/ESA Hubble Space Telescope, obtained 
at the Space Telescope Science Institute, which is operated by the Association of Universities 
for Research in Astronomy, Inc., under NASA contract NAS 5-26555. These observations are associated 
with program GO-10396.}
\altaffiltext{1}{Astronomical Institute, Department of Physics and Astronomy, 
University of Basel, Venusstrasse 7, CH-4102 Binningen, Switzerland}
\altaffiltext{2}{Department of Astronomy, University of Wisconsin, 475 North 
Charter Street, Madison, WI 53706-1582}
\altaffiltext{3}{Astronomisches Rechen-Institut, Zentrum f\"ur Astronomie der
Universit\"at Heidelberg, M\"onchhofstr.\ 12--14, D-69120 Heidelberg, Germany}
\altaffiltext{4}{Space Telescope Science Institute, 3700 San Martin Drive, 
Baltimore, MD 21218}
\altaffiltext{5}{INAF - Osservatorio Astronomico di Bologna, Via Ranzani 1, 
40127 Bologna, Italy}
\altaffiltext{6}{Department of Physics and Astronomy, University of California
at Los Angeles, 430 Portola Plaza, Los Angeles, CA 90095-1547}
\altaffiltext{7}{Research School of Astronomy \& Astrophysics, The Australian National 
University, Mt Stromlo Observatory, via Cotter Rd, Weston, ACT 2611, Australia}

\begin{abstract}
We present a photometric analysis of the star clusters Lindsay\,1, Kron\,3, NGC\,339, NGC\,416, Lindsay\,38,
and NGC\,419 in the Small Magellanic Cloud (SMC), observed with the \textit{Hubble Space Telescope} Advanced Camera for 
Surveys (ACS) in the F555W and F814W filters. Our color magnitude diagrams (CMDs) extend $\sim$3.5 mag deeper 
than the main-sequence turnoff points, deeper than any previous data. Cluster ages were derived using 
three different isochrone models: Padova, Teramo, and Dartmouth, which are all available in the ACS photometric system.
Fitting observed ridgelines for each cluster, we provide a homogeneous and unique set of low-metallicity, single-age 
fiducial isochrones. The cluster CMDs are best approximated by the Dartmouth isochrones for all clusters, except
for NGC\,419 where the Padova isochrones provided the best fit. Using Dartmouth isochrones we derive ages of 
$7.5 \pm 0.5$~Gyr 
(Lindsay\,1), $6.5 \pm 0.5$~Gyr (Kron\,3), $6 \pm 0.5$~Gyr (NGC\,339), $6 \pm 0.5$~Gyr (NGC\,416), and $6.5 \pm 0.5$~Gyr 
(Lindsay\,38). The CMD of NGC\,419 shows several main-sequence turn-offs, which belong to the cluster and to the SMC 
field. We thus derive an age range of 1.2-1.6~Gyr for NGC\,419. We confirm that the SMC contains several intermediate-age 
populous star clusters with ages unlike those of the Large Magellanic Cloud (LMC) and the Milky Way (MW). 
Interestingly, our intermediate-age star clusters 
have a metallicity spread of $\sim$0.6~dex, which demonstrates that the SMC does not have a smooth, monotonic 
age-metallicity relation. We find an indication for centrally concentrated blue straggler star candidates in 
NGC\,416, while for the other clusters these are not present. 
Using the red clump magnitudes, we find that the closest cluster, NGC\,419 ($\sim$50~kpc), and the 
farthest cluster, Lindsay\,38 ($\sim$67~kpc), have a relative distance of $\sim$17~kpc, which confirms the
large depth of the SMC. The three oldest SMC
clusters (NGC\,121, Lindsay\,1, Kron\,3) lie in the north-western part of the SMC, while the youngest (NGC\,419) 
is located near the SMC main body. 
\end{abstract}


\section{Introduction}
Star clusters are powerful tools for probing the star-formation history and the associated chemical evolution of a galaxy. 
As one of the closest star forming galaxies with star clusters covering a wide range of ages, the Small Magellanic Cloud 
(SMC) is a preferred location for detailed studies
of this class of objects. The SMC is the only dwarf galaxy in the Local Group containing populous intermediate-age star 
clusters of all ages. 
The SMC appears to be part of a triple system together with the Large Magellanic Cloud (LMC) and the Milky Way (MW). Its 
star formation activity may be triggered by interactions with its companions \citep[e.g.][]{yoshi03}. 
The proximity of the SMC allows us to resolve individual stars in compact and massive star clusters of intermediate 
and old age, down to the sub-solar stellar mass regime. 

The globular cluster (GC) system of the MW exhibits a range of ages between $\sim$10.5 and 14~Gyr \citep[e.g.,][]{deang05} 
with the oldest populations
belonging to the most ancient surviving stellar systems. In the Galactic halo, a ''young'' group of star clusters
is found with Pal\,1 being the youngest with an age of $8 \pm 2$~Gyr \citep{Rosenberg98}. Theories explaining the origin 
of these so-called young halo clusters, consider them to have been captured by the MW \citep{buon95}, to have been 
formed during 
interactions between the MW and the Magellanic Clouds \citep{fupe95}, or to have been accreted from destroyed and/or merged 
dwarf satellites \citep[e.g.,][]{zinn93,mack04}.

The star formation history of the LMC shows pronounced peaks that coincide with the times of possible past close 
encounters between the LMC, SMC and MW, indicative of interaction-triggered cluster formation \citep[e.g.,][]{gir95}. 
In the LMC, two epochs of cluster
formation have been observed that are separated by an ''age gap'' of about 4-9~Gyr \citep[e.g.,][]{holtz99, john99, Harris01}. 
In the early epoch a well-established population of metal-poor ($\langle[Fe/H]\rangle\sim -2$) star 
clusters with comparable properties to Galactic halo clusters \citep{sunt92, olsen98, dutra99} was 
formed. These clusters are as old as the oldest globular clusters in the MW and in the Galactic dwarf spheroidal companions
\citep{greb04}. In a second epoch, a large population of intermediate-age clusters with ages less than 3-4~Gyr 
have developed. 

In contrast, the SMC contains only one old GC, NGC\,121, which is 2-3~Gyr younger than the oldest GC 
in the LMC and MW \citep{glatt08} (Paper~I). The second oldest SMC star cluster, Lindsay\,1, has an age of 
$7.5 \pm 0.5$~Gyr, and since then compact populous star clusters have formed fairly continuously until the 
present day \citep[e.g.,][]{daco02}. Furthermore, the 
intermediate-age clusters in the SMC might survive for a Hubble time, due to their high mass and the structure of the SMC
(no bulge or disk to be passed) \citep{hunter03,lamers05,gieles07}.

The existing age determinations to this point have often been associated with large uncertainties. Stellar crowding, 
field star contamination and faintness of the main-sequence turnoffs made the measurement of precise ages difficult. 
These problems affect in particular ground-based data. 
Another difficulty is the large depth extent of the SMC which exacerbates the distance modulus-reddening
degeneracy for each cluster. These uncertainties can affect the age determination considerably. 

The capabilities of the Advanced Camera for Surveys (ACS) aboard the \textit{Hubble Space Telescope} (HST) provide an 
improvement both in sensitivity (depth) as well 
as angular resolution, which is essential for reliable photometric age determinations in dense clusters. 
We present improved cluster ages and distance determinations for Lindsay\,1, Kron\,3, NGC\,416, NGC\,339, Lindsay\,38, 
and NGC\,419. This is part of a ground-based and space-based program to uncover the age-metallicity evolution of the 
SMC. Our space-based imaging data were obtained with HST/ACS and our ground-based spectroscopy
was obtained with \textit{Very Large Telescope} (VLT). We combine our photometric results with spectroscopic metallicity 
determinations to obtain a well-sampled age-metallicity relation. 

The age-metallicity relation determined so far indicated that SMC clusters of similar age may differ by several 
tenths of dex in metallicity \citep{daco98}. Previous studies provided ages and metallicities of SMC star clusters 
using a variety of techniques and telescopes (see $\S$~\ref{sec:ana}). Combining all published cluster ages for e.g. 
Kron\,3 (5-10~Gyr) \citep{gasc66,alc96,migh98,udal98,rich00}, we find a wide range of ages for some key star clusters 
depending on the method used for the determination.

\begin{deluxetable*}{cccccc}
\tabletypesize{\scriptsize}
\tablecolumns{6}
\tablewidth{0pc}
\tablecaption{Journal of Observations}
\tablehead{
\colhead{Cluster} & \colhead{Date} & \colhead{} & \colhead{Total Exposure Time} & \colhead{R.A.} & \colhead{Dec.} \\
\colhead{} & \colhead{yy/mm/dd} & \colhead{Filter} & \colhead{(s)} & \colhead{} & \colhead{} } 
\startdata
Lindsay\,1     &2005$/$08$/$21& F555W & 40.0   & $0^h03^m53.19^s$ & $-73\arcdeg28'15.74''$ \\
	       &&       & 1984.0 & $0^h03^m52.66^s$ & $-73\arcdeg28'16.47''$\\
	       && F814W & 20.0   & $0^h03^m53.19^s$ & $-73\arcdeg28'15.74''$\\
	       &&	& 1896.0 & $0^h03^m52.66^s$ & $-73\arcdeg28'16.47''$\\
Kron\,3        &2006$/$01$/$17& F555W & 40.0   & $0^h24^m41.64^s$ & $-72\arcdeg47'47.49''$\\
	       &&	& 1984.0 & $0^h24^m41.92^s$ & $-72\arcdeg47'45.49''$\\
	       && F814W & 20.0   & $0^h24^m41.64^s$ & $-72\arcdeg47'47.49''$\\
	       &&	& 1896.0 & $0^h24^m41.92^s$ & $-72\arcdeg47'45.49''$\\
NGC\,339       &2005$/$11$/$28& F555W & 40.0   & $0^h57^m47.40^s$ & $-74\arcdeg28'26.25''$\\
	       &&	& 1984.0 & $0^h57^m47.13^s$ & $-74\arcdeg28'24.16''$\\
	       && F814W & 20.0   & $0^h57^m47.40^s$ & $-74\arcdeg28'26.25''$\\
	       &&	& 1896.0 & $0^h57^m47.13^s$ & $-74\arcdeg28'24.16''$\\
NGC\,416       &2006$/$03$/$08& F555W & 40.0   & $1^h07^m53.59^s$ & $-72\arcdeg21'02.47''$\\
(WFC)	       &&	& 1984.0 & $1^h07^m54.09^s$ & $-72\arcdeg21'01.79''$\\
	       && F814W & 20.0   & $1^h07^m53.59^s$ & $-72\arcdeg21'02.47''$\\
	       &&	& 1896.0 & $1^h07^m54.09^s$ & $-72\arcdeg21'01.79''$\\
NGC\,416       &2006$/$08$/$12& F555W & 70.0   & $1^h07^m58.76^s$ & $-72\arcdeg21'19.70''$\\
(HRC)	       &&	& 1200.0 & $1^h07^m58.96^s$ & $-72\arcdeg21'19.30''$\\
	       && F814W & 40.0   & $1^h07^m58.76^s$ & $-72\arcdeg21'19.70''$\\
	       &&	& 1036.0 & $1^h07^m58.96^s$ & $-72\arcdeg21'19.30''$\\        
Lindsay\,38    &2005$/$08$/$18& F555W & 40.0   & $0^h48^m57.14^s$ & $-69\arcdeg52'01.77''$\\
	       &&	& 1940.0 & $0^h48^m56.76^s$ & $-69\arcdeg52'03.07''$\\
	       && F814W & 20.0   & $0^h48^m57.14^s$ & $-69\arcdeg52'01.76''$\\
	       &&	& 1852.0 & $0^h48^m56.76^s$ & $-69\arcdeg52'03.07''$\\
NGC\,419       &2006$/$01$/$05& F555W & 40.0   & $1^h08^m12.53^s$ & $-72\arcdeg53'17.72''$\\
(WFC)	       &&	& 1984.0 & $1^h08^m12.71^s$ & $-72\arcdeg53'15.49''$\\
	       && F814W & 20.0   & $1^h08^m12.53^s$ & $-72\arcdeg53'17.72''$\\
	      &&       & 1896.0 & $1^h08^m12.71^s$ & $-72\arcdeg53'15.49''$\\
NGC\,419       &2006$/$04$/$26& F555W & 70.0   & $1^h08^m17.93^s$ & $-72\arcdeg53'03.60''$\\
(HRC)	       &&	& 1200.0 & $1^h08^m17.78^s$ & $-72\arcdeg53'02.80''$\\
	       && F814W & 40.0   & $1^h08^m17.93^s$ & $-72\arcdeg53'03.60''$\\
	      &&       & 1036.0 & $1^h08^m17.78^s$ & $-72\arcdeg53'08.80''$\\
\enddata
\label{tab:journalobs}
\end{deluxetable*}

Here we present the deepest available photometry with HST/ACS, which allows us to carry out the most accurate 
age measurements obtained so far. We determine the ages of these clusters utilizing three different isochrone 
models, which also yields distances. In the next Section we describe 
the data reduction procedure. In $\S$~\ref{sec:method} we present the color-magnitude diagrams (CMD) of the 
clusters and discuss their main features. In $\S$~\ref{sec:age} we describe our age derivation method and present 
our results. We give an estimate of the distances of our clusters long the line-of-sight in $\S$~\ref{sec:dist}
and present a discussion and a summary in Sections $\S$~\ref{sec:ana} and $\S$~\ref{sec:sum}, respectively.

\begin{figure*}
  \epsscale{1.2}
  \plotone{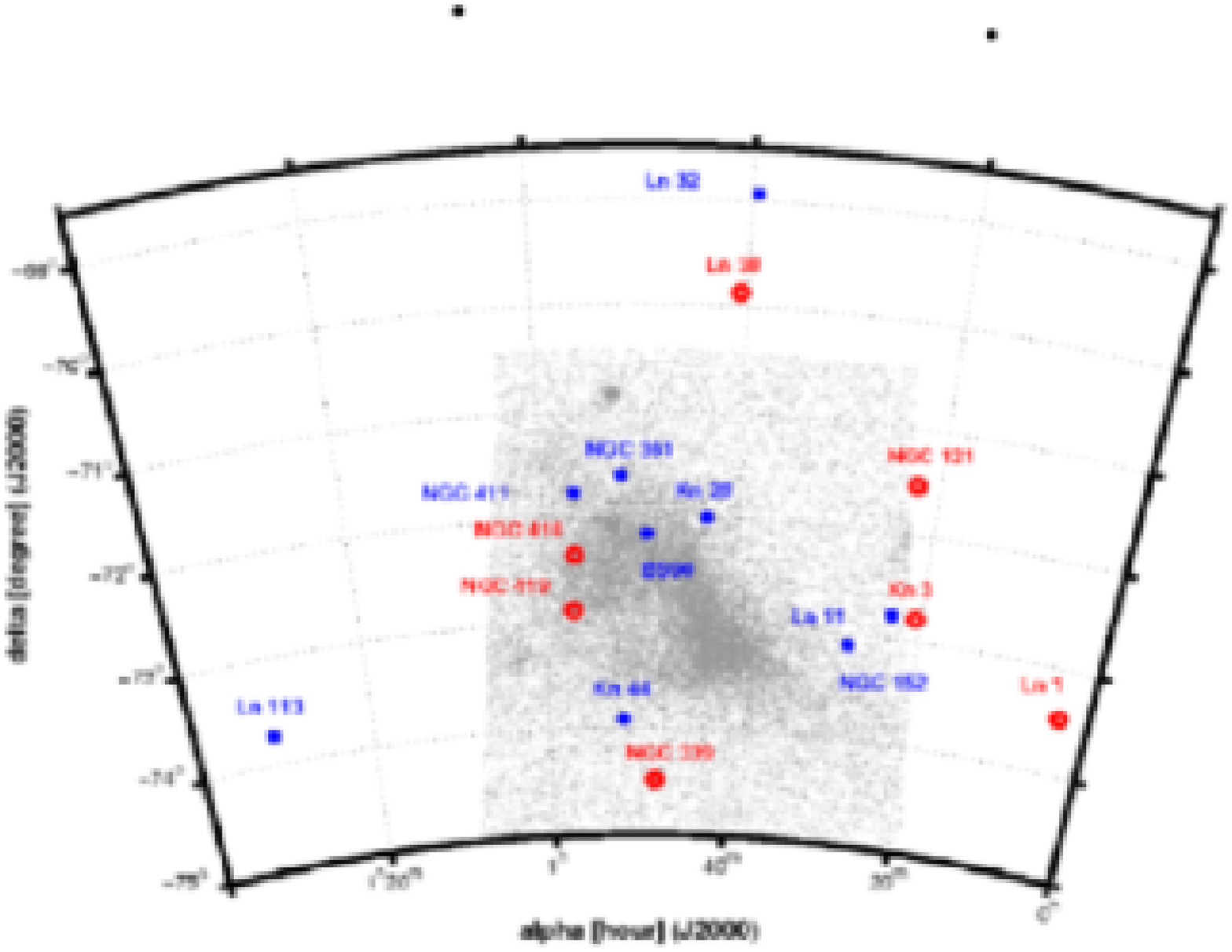}
  \caption{Spatial Distribution in 2D of our cluster sample (red circles). The location of eight additional SMC 
  clusters, for which reliable ages from the literature are available, is shown (blue crosses). We obtain 
  a complete sample of all intermediate-age and old SMC star clusters (see $\S$~\ref{sec:age}), which we will 
  discuss in $\S$~\ref{sec:dist} and~\ref{sec:ana}. One of the clusters, Lindsay\,116, lies outside the coordinate 
  boundaries of the Figure. The cluster locations are shown superimposed on a star map of the SMC 
  generated using the point source catalog of the Small Magellanic Cloud Photometric Survey \citep{zar02} for 
  stars with V$<$16.5~mag.}
  \label{fig:clusters}
\end{figure*}

\section{Observations and Reductions}

The SMC clusters Lindsay\,1, Kron\,3, NGC\,339, NGC\,416, Lindsay\,38, and NGC\,419 were observed with the 
HST/ACS between 2005 August and 2006 March 
(Table~\ref{tab:journalobs}). The observations are part of a project (GO-10396; principal investigator: 
J.~S.~Gallagher, III) that is focused on cluster and field populations and the star formation history of 
the SMC. 

The images were taken with the F555W and F814W filters, which closely resemble the Johnson V and I filters in 
their photometric properties \citep{siri05}. For Lindsay\,1, Kron\,3, NGC\,416, NGC\,339, and Lindsay\,38 we 
discuss photometry from the \textit{Wide Field Camera} (WFC), while for NGC\,419 and for the center region of 
NGC\,416 photometry from the \textit{High Resolution Camera} (HRC) was used. The WFC images cover an area of 
$200'' \times 200''$ at each pointing with a pixel scale of $\sim$0.05~arcsec. The HRC images cover an 
area of $29'' \times 26''$ at each pointing with a pixel scale of $\sim$0.025~arcsec.

The data sets were processed adopting the standard Space Telescope Science Institute ACS calibration pipeline 
(CALACS) to subtract the bias level and to apply the flat field correction. For each filter, the short and 
long exposures were co-added independently using the MULTIDRIZZLE package \citep{koek02}. Cosmic rays and hot 
pixels were removed with this package and a correction for geometrical distortion was provided. Because
these mainly affect the faint stars we did not perform CTE corrections. The resulting 
data consist of one 40~s and one 1984~s exposure (1940~s for Lindsay\,38) in F555W and one 20~s as well as one 
1896~s exposure (1852~s for Lindsay\,38) in F814W (Tab.~\ref{tab:journalobs}). The HRC data of NGC\,419 
consist of 70~s and 1200~s exposure in F555W and 40~s and 1036~s exposure in F814W each.

The photometric reductions were carried out using the DAOPHOT package in the IRAF 
\footnote{\scriptsize{\small{IRAF} is 
written and supported by the IRAF programming group at the National Optical Astronomy Observatories (NOAO) 
in Tuscon, Arizona. NOAO is operated by the Association of Universities for Research in Astronomy, Inc. 
under cooperative agreement with the National Science Foundation.}} environment on DRIZZLed images. 

\textit{WFC}: Saturated foreground stars and background galaxies were discarded by using the Source Extractor 
\citep{bert96}. Due to the different crowding and signal-to-noise ratio properties of the long and the short 
exposure, photometry involving point spread function (PSF) fitting was only performed on the long exposures. 
For the short exposures, we used aperture photometry, which turned out to yield smaller formal errors than PSF 
photometry (see also Paper~I). 
The detection thresholds were set at 3~$\sigma$ above the local background level for Lindsay\,1, 1~$\sigma$ for 
Kron\,3 and 4~$\sigma$ for NGC\,339, NGC\,416, and Lindsay\,38 in order to detect even the faintest 
sources. The threshold levels were set based on the different crowding effects of the single clusters. 
The data reduction and photometry for the WFC images followed exactly the procedures outlined in 
Paper~I, and for a detailed description we refer the reader to this paper.

\textit{HRC (NGC\,416, NGC\,419)}: The photometry was carried out independently in F555W and F814W.
The detection threshold was set at 4~$\sigma$ for both NGC\,416 and NGC\,419 above the local 
background level. 

For those stars common in both filters, we first performed aperture photometry using an aperture 
radius of 2 pixels to avoid the diffraction ring. Due to the fact that the PSF does not vary on the HRC images, a 
PSF then was constructed by combining 10 bright and isolated stars that were uniformly distributed over the entire 
image. The PSF photometry was then carried out. The objects found in both images were cross-identified and merged 
with a software package written by P.~Montegriffo (private communication). 

A spatial projection of the clusters' location towards the SMC (\textit{red circles}) is shown in Figure~\ref{fig:clusters} 
superimposed on a star map of the SMC generated using the point source catalog of the Small Magellanic Cloud 
Photometric Survey \citep{zar02} for stars V$<$16.5~mag. Additionally, we show the location of the eight 
intermediate-age star clusters (\textit{blue crosses}), which we will discuss in $\S$~\ref{sec:ana}.

\begin{figure}
  \epsscale{1}
  \plotone{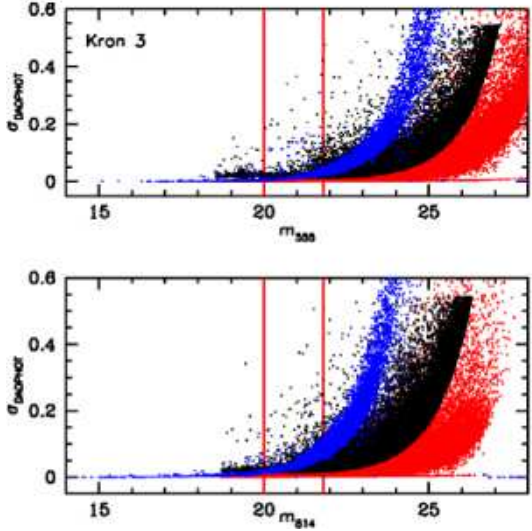}
  \caption{Photometric errors assigned by DAOPHOT to stars on the short (blue dots), and on the long (aperture 
  photometry: red dots, PSF photometry: black dots) exposures for the cluster Kron\,3. Stars brighter than 
  $\sim$18.5~mag in the long F555W exposure and brighter than $\sim$18.7~mag in the long F814W exposure are 
  saturated and are therefore not shown. The samples of the short, and the long exposures measured with aperture 
  photometry and PSF photometry are combined at 21.8~mag and $m_{555}$ = 20~mag (indicated by the thin vertical 
  lines). For the F814W exposures we chose the same magnitude value in order to avoid introducing a color slope 
  in the color-magnitude diagram of the resultant data set.}
\label{fig:k3_sigma}
\end{figure}

We show the photometric errors assigned by DAOPHOT for Kron\,3 in Figure~\ref{fig:k3_sigma} as these are representative
of our WFC data. The formal photometric errors remain negligible over a wide range of magnitudes for 
stars measured on the short exposures. Photometry obtained with aperture photometry on the long exposure yields 
smaller errors for stars brighter than $m_{555,814} \sim 22$~mag than with PSF photometry. In Kron\,3, all 
brighter stars in the long exposures down to $m_{555} < 18.5$~mag and $m_{814} < 18.7$~mag are saturated.
For stars brighter than $m_{555,814} = 20$~mag, the short exposure sample (\textit{blue dots}), in the interval
between $21.8 <  m_{555,814} < 20$~mag the long exposure aperture photometry sample (\textit{red dots}) and for 
stars fainter than $m_{555,814} = 21.8$~mag the long exposure PSF photometry sample (\textit{black dots}) was used. 
We determined the cuts between the samples based on the $m_{555}$ data and adopted the same value 
for $m_{814}$ so as to avoid a color slope associated with this division. 

For our study, we rejected all stars with a $\sigma$ error larger than 0.2~mag and a DAOPHOT sharpness parameter 
$-0.2 \leq s \leq 0.2 $ in both WFC and HRC filters. The resulting color-magnitude diagrams (CMD) of Lindsay\,1, 
Kron\,3, NGC\,339, NGC\,416, Lindsay\,38, and NGC\,419 are shown and discussed in the following Section.

\section{THE COLOR-MAGNITUDE DIAGRAMS}
\label{sec:method}

All the CMDs of our six clusters show a well-populated main-sequence (MS), and clearly defined sub-giant 
and red-giant branches (SGB and RGB, respectively). The asymptotic giant branch (AGB) is less tightly defined,
but clearly present in all cases, and especially evident in NGC\,416 (Fig.~\ref{fig:ngc416_errorbar_ap}).
We define the main-sequence turnoff-point (MSTO) to represent the bluest point on the 
observed ridgeline. The data allow us to carry out the most accurate age measurement obtained so far (see 
Section~\ref{sec:age}), while also deriving improved distances. Lindsay\,1, Kron\,3, NGC\,339, NGC\,416, and 
Lindsay\,38 appear to be single-age, simple stellar  
population objects, while the WFC data of NGC\,419 shows multiple turnoff-points. For this reason we will discuss 
this cluster in greater detail in a separate paper (Sabbi et al. 2008, in preparation). 

The CMDs show no obvious evidence for Galactic foreground contamination due to the high Galactic latitude of 
the SMC \citep[e.g., ][]{ratna85}. However, we find significant SMC field star contamination in the 
CMDs of Kron\,3, NGC\,339, NGC\,416, and also in the HRC CMD of NGC\,419. Field stars naturally are more problematic in 
CMDs of clusters near the SMC main body. For 
all clusters we give the magnitude of the MSTO, the red clump, and the red bump in Table~\ref{tab:prop}. 
For the red clump and the red bump the mean magnitude was calculated by averaging the magnitudes of 
all clump and bump stars respectively and finding the maximum of each luminosity function. 

For Kron\,3 and Lindsay\,38 no red bump was found. The red bump is a feature predicted by stellar evolution models, 
which imply that the luminosity of the RGB bump is dependent on the metallicity and age of the cluster. 
The failure to identify a red bump in the CMD of Lindsay\,38 is due to its sparseness. The CMD of Kron\,3,
however, is well-populated, but despite sufficient statistics a red bump is not present, for which we have
no physical explanation. In the 
CMDs of Lindsay\,1, Kron\,3, and NGC\,339 a gap on the RGB at $\sim$20~mag is visible. For Lindsay\,1
and Kron\,3 this feature is artificial due to the cuts in photometric errors and sharpness we applied. 
However, for NGC\,339 the gap appears to be real (see $\S$~\ref{sec:ngc339_cmd}). 

\subsection{Lindsay\,1}
\label{sec:L1_cmd}

\begin{figure}
  \epsscale{1}
  \plotone{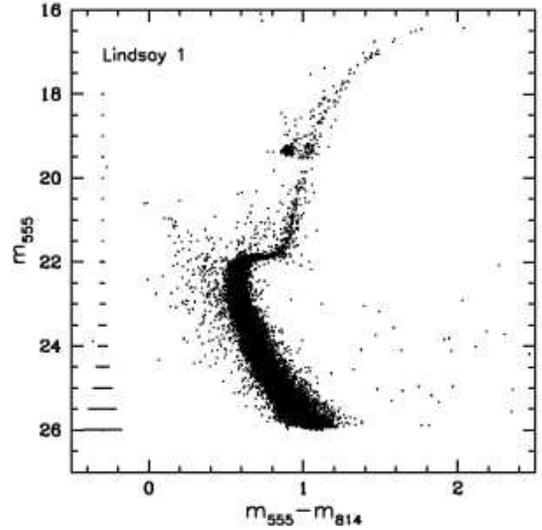}
  \caption{CMD of Lindsay\,1 and its surroundings. Only stars with good photometry ($\sigma \leq 0.2$~mag and 
  $-0.2 \leq$ sharpness $\leq 0.2$) are shown; 15,321 stars in total. Representative errorbars (based on errors 
  assigned by DAOPHOT) are shown on the left for the $m_{555}$-$m_{814}$ color.}
  \label{fig:l1_errorbar_ap}
\end{figure}

The populous cluster Lindsay\,1 \citep{lind58} is the westernmost known cluster in the SMC and lies around 
3.5\degr~west of the bar. Lindsay\,1 is the second oldest star cluster in the SMC after NGC~121 (e.g., Paper~I). 
The color-magnitude diagram (CMD) of Lindsay\,1 is presented in Figure~\ref{fig:l1_errorbar_ap}. 
The CMD reaches $\sim$3~mag deeper than the previous deepest available photometry \citep{migh98,alc03}. 
Lindsay\,1 is located in a low-density area outside the main body of the SMC (see Fig.~\ref{fig:clusters}). 
Therefore we find only little field star contamination by younger populations near Lindsay\,1. 
 
\begin{figure}
  \epsscale{1}
  \plotone{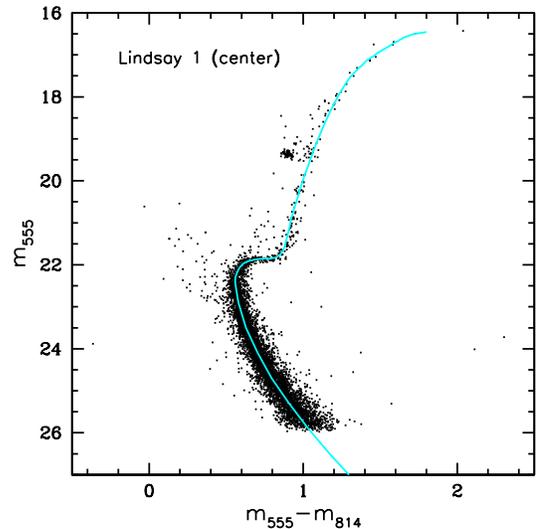}
  \caption{CMD of the central region of Lindsay\,1. All stars within a radius of 45'' were selected. We used 
  this CMD for the determination of a representative color-magnitude ridgeline (cyan line) of Lindsay\,1. This CMD contains 
  5,561 stars.}
\label{fig:l1_ridgeline}
\end{figure}

In Figure~\ref{fig:l1_ridgeline} we display the CMD for the stars in the center region of Lindsay\,1. The location 
of the cluster center was visually estimated. All stars within a radius of 45'' were selected to create the center 
sample. Most of the field contamination has vanished. Given the width of our MS, we cannot infer any
information on the presence and percentage of unresolved binary systems.

We find a well-populated red clump at a mean magnitude of $m_{555}$ = $19.36 \pm 0.04$~mag. Our value is in 
excellent agreement with the horizontal branch magnitude found by 
\citet[][and Alcaino et al.(2003)]{sara95,alsa99,rich00,crowl01}. Obviously, Lindsay\,1 is not old enough to have 
developed an extended red horizontal branch, which in itself is already a strong indication that Lindsay\,1 is 
younger than NGC\,121. We refer to \citet{salgir02} who study the behavior of the red clump as a function of age. 

The red clump of Lindsay\,1 is $\sim$0.35~mag brighter in F555W than that of NGC\,121. The 
luminosity difference may imply that Lindsay\,1 is closer than NGC\,121 along the line-of-sight, 
which actually seems to be the case (see $\S$~\ref{sec:dist}). Adopting the absolute red clump magnitudes given 
by \citet{gir01}, and the reddening values from the Schlegel maps \citep{Schlegel98} of $E_{B-V} = 0.03$~mag, 
we find an absolute red clump magnitude difference between NGC\,121 and Lindsay\,1 of 
$\Delta M^{RC}_{m555} \sim$ 0.28~mag. Therefore, the feature seen in Lindsay\,1, should be considered a red clump 
and not a red horizontal branch.

The gap on the RGB at $m_{555} \sim$ 19.8~mag is an artificial feature due to small number statistics resulting 
from our photometric selection (only photometry with $\sigma <$ 0.2~mag and $-0.2 \leq s \leq 0.2 $ is shown). 

\subsection{Kron\,3} 
\label{sec:k3_cmd}
 
 \begin{figure}
  \epsscale{1}
  \plotone{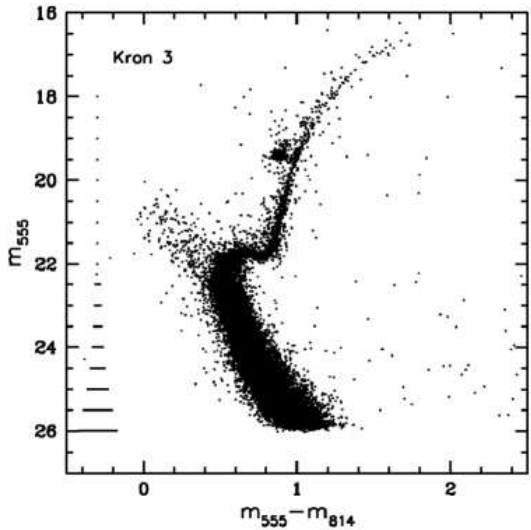}
  \caption{CMD of Kron\,3 and its surroundings. Only stars with good photometry ($\sigma \leq 0.2$~mag and 
  $-0.2 \leq$ sharpness $\leq 0.2$) are shown; 30,264 stars in total. Representative errorbars (based on errors 
  assigned by DAOPHOT) are shown on the left for the $m_{555}$-$m_{814}$ color.}
  \label{fig:k3_errorbar_ap}
\end{figure}

\begin{figure}
  \epsscale{1}
  \plotone{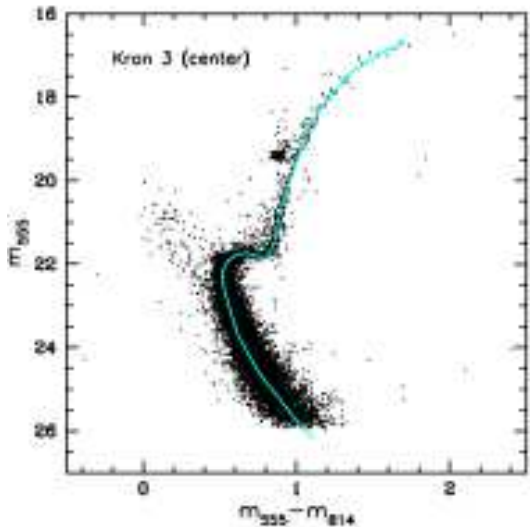}
  \caption{CMD of the central region of Kron\,3. All stars within a radius of 40'' were selected. We used 
  this CMD for the determination of a representative color-magnitude ridgeline (cyan line) of Kron\,3. This CMD contains 
  13,584 stars.}
\label{fig:k3_ridgeline}
\end{figure}

Kron\,3 lies well outside the main SMC body, about 2\degr~west of the bar. The cluster was first cataloged by 
\citet{shapwil25}, and it is number 3 in Kron's (1956) catalog of SMC star clusters. The highly populated 
CMD of Kron\,3 is presented in Figure~\ref{fig:k3_errorbar_ap}. The CMD reaches $\sim$2 magnitudes deeper 
than the previous deepest available photometry \citep{rich00}. 

Field star contamination is visible along an extension of the main-sequence, towards brighter and bluer 
objects than at the cluster's MSTO. However, the cluster center is not affected by crowding, even though 
the center region is very dense. 

In Figure~\ref{fig:k3_ridgeline} we display the CMD for the stars within 40'' of the center of Kron\,3. 
From the width of the MS, we cannot draw any conclusions about the presence of unresolved binary systems. 
The aforementioned traces of field contamination are still visible.

\subsection{NGC\,339}
\label{sec:ngc339_cmd}

\begin{figure}
  \epsscale{1}
  \plotone{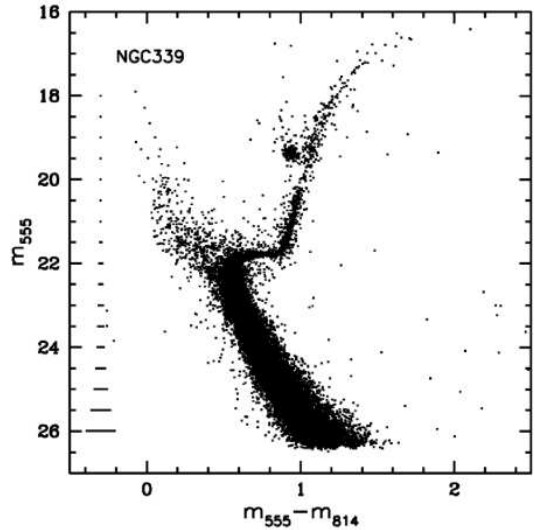}
  \caption{CMD of NGC\,339 and its surroundings. Only stars with good photometry ($\sigma \leq 0.2$~mag and 
  $-0.2 \leq$ sharpness $\leq 0.2$) are shown; 29,304 stars in total. Representative errorbars (based on errors 
  assigned by DAOPHOT) are shown on the left for the $m_{555}$-$m_{814}$ color.}
  \label{fig:ngc339_errorbar_ap}
\end{figure}

\begin{figure}
  \epsscale{1}
  \plotone{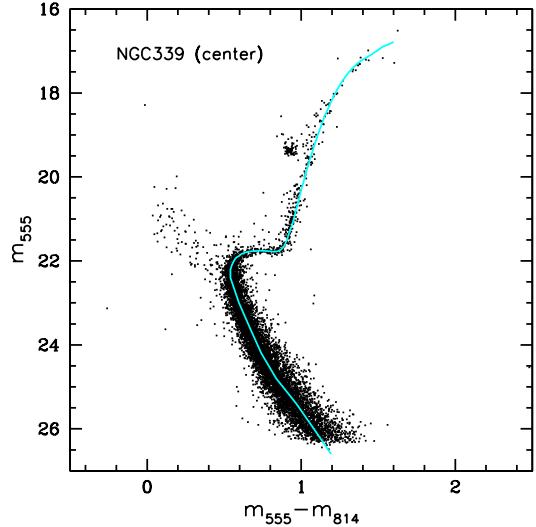}
  \caption{CMD of the central region of NGC\,339. All stars within a radius of 35'' were selected. We used 
  this CMD for the determination of a representative color-magnitude ridgeline (cyan line) of NGC\,339. This CMD contains 
  8,555 stars.}
\label{fig:ngc339_ridgeline}
\end{figure}

NGC\,339 is located outside the SMC main body, around 1\degr~south of the bar. The resulting CMD is shown in 
Figure~\ref{fig:ngc339_errorbar_ap}, which reaches $\sim$2 magnitudes deeper than the previous deepest available 
photometry, published by \citet{rich00}.

The SMC field is present along the luminous extension of the main-sequence. The cluster center is fully 
resolved and not affected by crowding since this is a very low density cluster. All features in our CMD are well 
defined. We cannot infer any information on unresolved binary systems due to the width of our MS. Unlike for
the other clusters in our sample, the gap on 
the RGB at $m_{555} \sim$ 20~mag is not an artificial feature. It is visible in both the single short and long
exposures and has also been found in other SMC clusters, e.g. NGC\,288 \citep{Bellazini01}.

To create the center sample, all stars within a radius of 35'' around a visually estimated center were selected and 
displayed in Figure~\ref{fig:ngc339_ridgeline}. The SMC field is still clearly visible, which was expected due
to the location of the cluster close to the SMC main body.

\subsection{NGC\,416}
\label{sec:ngc416_cmd}

\begin{figure}
  \epsscale{1}
  \plotone{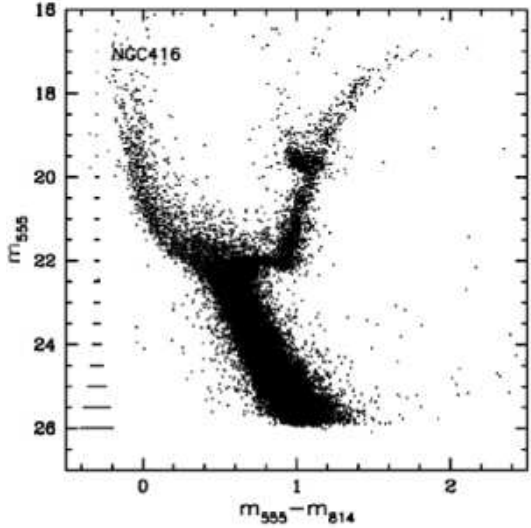}
  \caption{CMD of NGC\,416 and its surroundings. Only stars with good photometry ($\sigma \leq 0.2$~mag and 
  $-0.15 \leq$ sharpness $\leq 0.15$) are shown; 18,764 stars in total. Additionally, we discarded all stars 
  located within a radius of 15'' around the cluster center, due to the high density of the cluster center and 
  the resulting insufficient photometry. Representative errorbars (based on errors assigned by DAOPHOT) are shown 
  on the left for the $m_{555}$-$m_{814}$ color.}
\label{fig:ngc416_errorbar_ap}
\end{figure}

\begin{figure}
  \epsscale{1}
  \plotone{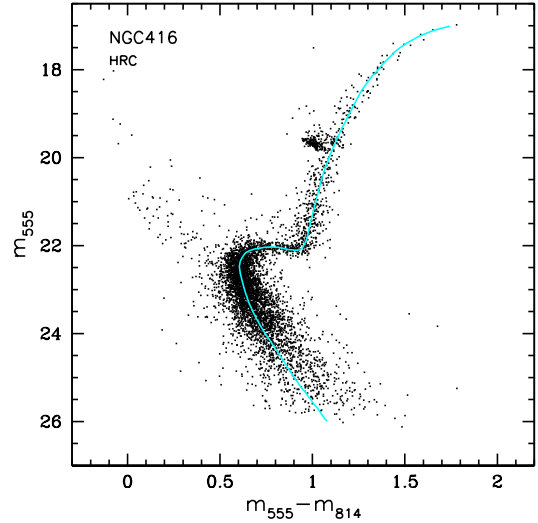}
  \caption{CMD of the HRC data of the center region of NGC\,416. We used 
  this CMD for the determination of a representative color-magnitude ridgeline (cyan line) of NGC\,416. This CMD contains 
  4,992 stars.}
\label{fig:ngc416_ridgeline}
\end{figure}

The cluster NGC\,416 is located in the wing of the SMC. This part of the SMC is characterized by an increased 
stellar density that may represent a tidal extension towards the LMC. Due to the location of NGC\,416 in 
the wing of the SMC, we expect a very rich CMD with strong SMC field star features. Our resulting CMD 
(Fig.~\ref{fig:ngc416_errorbar_ap}) indeed presents a densely populated MS, SGB, RGB, AGB and red clump as well as 
more luminous blue MS and blue loop stars stars that belong to younger SMC field populations.
The RGB is also broadened by SMC field stars and is not as narrow as in the other clusters in our sample.
About 0.1~mag offset to the blue of the RGB, we find a very well-populated AGB.
 
To obtain the final CMD we combine the HRC and the WFC photometry and discard the overlapping center region 
from the WFC catalog. For the WFC catalog, we only use PSF photometry of the long exposures and aperture 
photometry of the short exposures, because aperture photometry could not resolve single stars in the dense 
center region. The high density of the cluster center 
made it very difficult to find bright and isolated stars for the PSF sample in the cluster center. For the HRC 
data the long exposure images provided the better photometric quality. Therefore we do not include the short 
exposures in our final HRC catalog.

In Figure~\ref{fig:ngc416_ridgeline} we display only the HRC data of the center region of NGC\,416, which was 
used for the ridgeline fit. The CMD is still densely populated and all cluster features are clearly outlined. 
Interestingly, the RGB of NGC\,416 is still broadened by contaminating field stars. The main-sequence is well 
defined until $\sim$24~mag and fades out for fainter magnitudes. A number of younger stars of the SMC field are still 
visible above the MSTO. 

\subsection{Lindsay\,38}
\label{sec:l38_cmd}

\begin{figure}
  \epsscale{1}
  \plotone{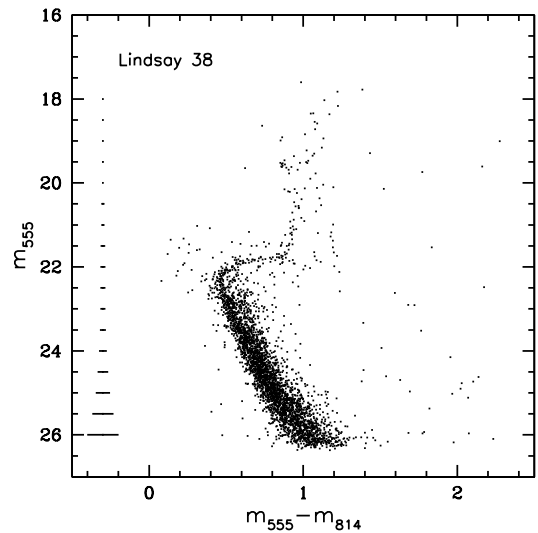}
  \caption{CMD of Lindsay\,38 and its surroundings. Only stars with good photometry ($\sigma \leq 0.2$~mag and 
  $-0.2 \leq$ sharpness $\leq 0.2$) are shown; 3,716 stars in total. Representative errorbars (based on errors 
  assigned by DAOPHOT) are shown on the left for the $m_{555}$-$m_{814}$ color.}
  \label{fig:l38_errorbar_ap}
\end{figure}

\begin{figure}
  \epsscale{1}
  \plotone{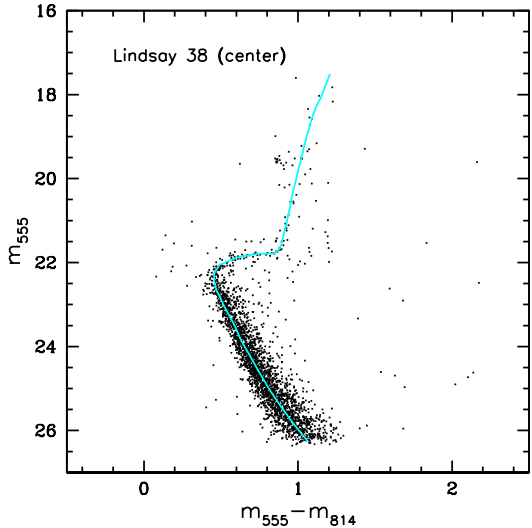}
  \caption{CMD of the central region of Lindsay\,38. All stars within a radius of 70'' were selected. We used 
  this CMD for the determination of a representative color-magnitude ridgeline (cyan line) of Lindsay\,38. This CMD contains 
  1,151 stars.}
  \label{fig:l38_ridgeline}
\end{figure}

Lindsay\,38 is located about 3.3\degr~north of the bar and is among the outermost SMC clusters. \citet{pia01} 
published the first and as yet only CMD of Lindsay\,38. Our resulting CMD of the custer and its surroundings is
shown in 
Figure~\ref{fig:l38_errorbar_ap}. Our CMD reaches $\sim$3~mag deeper than the CMD presented by \citet{pia01}, which
was obtained from ground-based photometry. The cluster is very sparse and we identify only 3,716 candidate member
stars. 

In Figure~\ref{fig:l38_ridgeline} we show the CMD for the center sample. We selected stars within a radius of 70''
around a visually estimated center location. The radius is almost twice as large as for the other clusters, due
to the sparseness of the cluster.

The MS and the SGB are nevertheless well defined. Only a few stars populate the upper RGB, but the red clump is 
clearly visible. 

\subsection{NGC\,419}
\label{sec:ngc419_cmd}

Like NGC\,416, the cluster NGC\,419 is located in the wing of the SMC. For this cluster, we show in this paper 
only the HRC data, and will discuss and analyze the full CMD in a separate paper due to its complexity. The CMD
reaches $\sim$2~mag deeper than the previous deepest available photometry, published by \citet{rich00}.

\begin{figure}
  \epsscale{1}
  \plotone{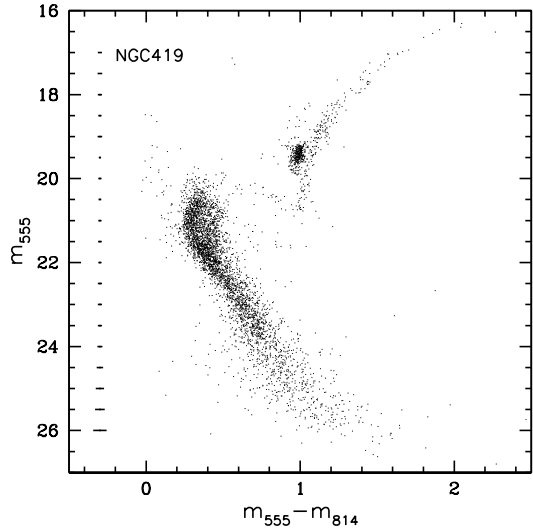}
  \caption{CMD of NGC\,419 from the HRC images. Only stars with good photometry ($\sigma \leq 0.2$~mag) 
  are shown; 4543 stars in total. Representative errorbars (based on errors assigned by DAOPHOT) are shown on the 
  left for the $m_{555}$-$m_{814}$ color.}
  \label{fig:n419_errorbar}
\end{figure}

Figure~\ref{fig:n419_errorbar} displays our CMD of NGC\,419, the youngest cluster in our sample. Only long exposure 
PSF photometry was used. The upper MS is rather broad and densely populated. The CMD reaches $\sim$4.5~mag below the 
MSTO, but due to shorter exposure times and crowding, the MS becomes less densely populated at fainter magnitudes. 
The tip of the extended turn-off region lies at $m_{555} \sim$20~mag, and more than one MSTO appear to be visible. 
For stellar populations in the corresponding age range ($\sim$1-1.6~Gyr), the hydrogen-shell burning phase lasts only 
for a short time ($\sim$70~Myr), which explains the sparse SGB \citep{Schaller92}. 

At $m_{555} \sim$21.5~mag and 
$m_{555}-m_{814} \sim 0.5-0.9$~mag in the CMD we find $\sim$30 SMC field stars, which may belong to an older MSTO.
\citet{mackey08} (see also Mackey $\&$ Broby Nielsen 2007) discovered the only two star clusters in the LMC 
known to have a double MS, NGC\,1846 and NGC\,1806, which both have similar metallicities of about Z = 0.0075. 
The CMDs look very similar to NGC\,419. Padova isochorones were used to determine the ages of NGC\,1846 and 
NGC\,1806 and yielded ages of 1.6 and 1.9~Gyr for both NGC\,1846 and NGC\,1806. 

\begin{figure}
  \epsscale{1}
  \plotone{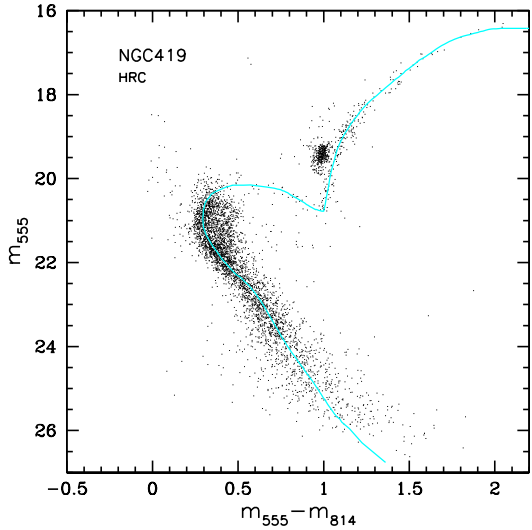}
  \caption{CMD of the central region of NGC\,419. The derived fiducial ridgeline is shown as the cyan line.}
 \label{fig:N419_ridgeline}
\end{figure}

In Figure~\ref{fig:N419_ridgeline} we show the CMD with the derived ridgeline. At $19.41 \pm 0.12$~mag we find 
the vertically extended red clump. Below the red clump, $\sim$0.2~mag fainter and $\sim$0.1~mag in color to the blue, we 
find parts of a second red clump population at $19.53 \pm 0.17$~mag. For the old globular cluster NGC\,121, we 
found the red clump at $19.71 \pm 0.03$~mag. Therefore we speculate that these fainter stars belong to a red clump 
of the old SMC field star population. 
If the luminosity difference (0.12~mag) between the two putative red clump populations were primarily due to 
distance (not age), then the distance d between the two populations can be calculated. We obtain $\delta$d = 4~pc. 

The MSTO is located at $21.4 \pm 0.1$~mag and $0.41 \pm 0.05$~mag. We have to be cautious with the reliability 
of the determined MSTO, because as we will see in $\S$~\ref{sec:agen419} the isochrone models describe a 
hook-like feature past the MSTO, which is not visible in our CMD. This phase is traversed quite rapid 
($\sim$27~Myr) for stars with ages between 1 and 1.6~Gyr \citep{Schaller92}. This short phase lifetime is likely 
the reason why the hook is not visible.

\begin{deluxetable}{cccc}
\tablecolumns{4}
\tablewidth{0pc}
\tablecaption{Observational data}
\tablehead{
\colhead{Cluster} & \colhead{$m_{555,TO}$} & \colhead{$m_{555,Bump}$}  & \colhead{$m_{555,RC}$}\\
\colhead{} & \colhead{mag} & \colhead{mag} & \colhead{mag} } 
\startdata
NGC\,121    & $22.98 \pm 0.05$ & $19.52 \pm 0.04$ & $19.71 \pm 0.03$   \\
Lindsay\,1  & $22.36 \pm 0.05$ & $19.25 \pm 0.05$ & $19.36 \pm 0.04$   \\
Kron\,3     & $22.40 \pm 0.05$ & -		& $19.46 \pm 0.03$   \\
NGC\,339    & $22.30 \pm 0.05$ & $19.27 \pm 0.04$ & $19.38 \pm 0.08$  \\
NGC\,416    & $22.44 \pm 0.05$ & $19.65 \pm 0.06$ & $19.70 \pm 0.07$  \\
Lindsay\,38 & $22.36 \pm 0.05$ & -		& $19.60 \pm 0.05$   \\
NGC\,419    & $21.40 \pm 0.10$ & $18.86 \pm 0.12$ & $19.41 \pm 0.12$   \\
\enddata
\label{tab:prop}
\end{deluxetable}

\section{BSS candidates}
\label{sec:bss}

\begin{figure}
  \epsscale{1}
  \plotone{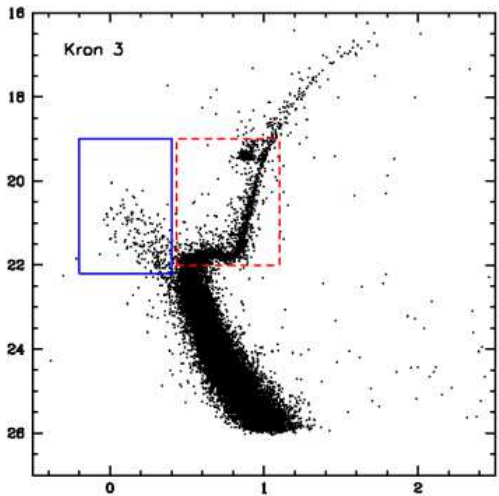}
  \caption{CMD of Kron\,3 with displayed sample selection for the BSS sample (solid lines) and the cluster 
  sample (dashed lines) including SGB, lower RGB, and red clump.}
  \label{fig:BSS_sample}
\end{figure}

\begin{figure}
  \epsscale{1.3}
  \plotone{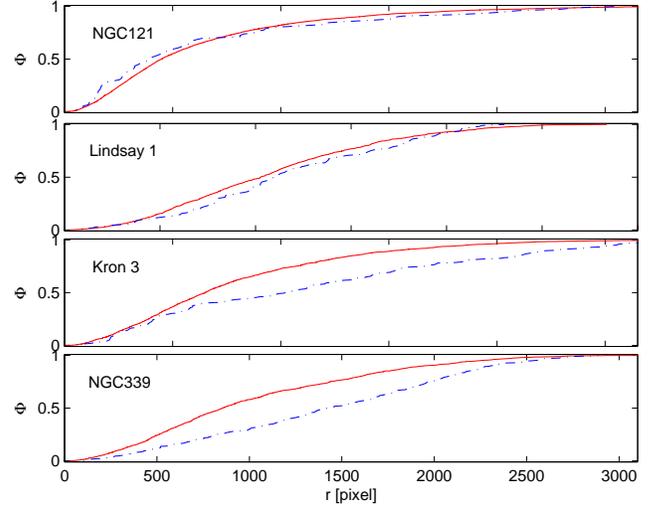}
  \caption{Cumulative radial distributions of blue straggler candidates as a function of projected radius for the clusters 
  NGC\,121, Lindsay\,1, Kron\,3, and NGC\,339. The solid line represents the 'cluster sample' including the SGB, 
  RGB, and red clump. The dashed line represents the stars found in the BSS region. The uppermost panel shows
  the distributions of NGC\,121 for comparison, because we know it contains blue straggler stars 
  (Shara et al. 1998, Clementini et al. in prep.). It is clearly visible that the BSS candidates are associated with 
  the cluster rather than the field population. For Lindsay\,1, Kron\,3, and NGC\,339, 
  the BSS candidates are not as clearly concentrated in the cluster center and are rather younger main-sequence objects of 
  the SMC field.}
  \label{fig:KS1}
\end{figure}

\begin{figure}
  \epsscale{1.3}
  \plotone{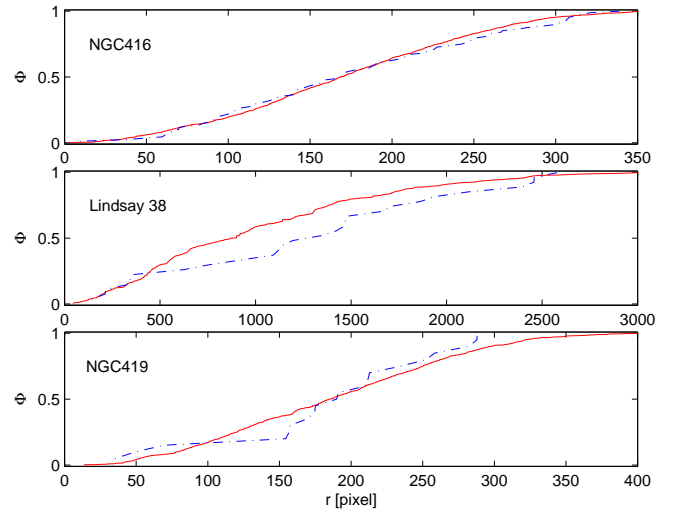}
  \caption{Cumulative radial distributions of blue straggler candidates with projected radius for the clusters NGC\,416, 
  Lindsay\,38, and NGC\,419. The solid line represents the 'cluster sample' including the SGB, RGB, and red clump. 
  The dashed line represents the stars found in the BSS region. In Lindsay\,38 and NGC\,419, the BSS candidates are not 
  significantly concentrated in the cluster center and are likely regular main-sequence stars. The distributions 
  of NGC\,416 and NGC\,419 are obtained using the HRC sample. In NGC\,416 there is indication for centrally 
  concentrated BSS candidates.}
  \label{fig:KS2}
\end{figure}

The stars blueward of and above the cluster MSTO are blue straggler star (BSS) candidates. In the same 
region often stars of the surrounding younger field population of the host galaxy are located. Knowledge 
of the BSS population of clusters is of interest with respect to constraining the binary fraction in these 
objects and with regard to understanding the formation mechanism of the BSS themselves 
\citep[e.g., ][]{bailyn95}. BSS have been detected in a wide range of cluster types including very young, 
populous Magellanic Cloud clusters \citep[e.g., ][]{greb96}, and old globular clusters 
\citep[e.g., ][]{Ferraro03}. \citet{carraro08} showed for three open Galactic clusters that BSS are centrally 
concentrated when comparing the CMDs of the clusters and the surrounding field as one would expect for 
populations associated with a cluster. These authors re-emphasized the problem of field star contamination 
when trying to photometrically identify BSS. 

Our ACS data do not cover fields adjacent to the clusters. Therefore, we calculated the projected distance from 
the cluster center for each BSS candidate. We then constructed a cumulative distribution of the number of 
blue stragglers as a function of projected radius. We have selected the BSS candidates by defining a region 
above the cluster main-sequence turnoff in the CMDs. We show the sample selection in the CMD of Kron\,3 in 
Figure~\ref{fig:BSS_sample} as a representative example.

The cumulative radial distributions of the selected stars for all clusters can be seen in Figures~\ref{fig:KS1}, 
and~\ref{fig:KS2}. The dashed lines show the cumulative distributions for our BSS candidate samples, and the 
solid lines show the cumulative distributions of the cluster sample for comparison, including the SGB, the 
lower RGB, and the red clump. 

Each BSS candidate was examined by eye on the image and stars with hints of being affected by blends were 
eliminated from the catalog (see Table~\ref{tab:BSS}). The cumulative distributions of the remaining stars 
are normalized to our BSS candidate sample. The first panel of Figure~\ref{fig:KS1} shows the cumulative 
distributions of NGC\,121, of which we know that it contains BSSs (Shara et al. 1998, Clementini et al. in 
preparation). The BSS candidates are evidently more centrally concentrated than the stars from the 
cluster sample. 

In Lindsay\,1, Kron\,3, NGC\,339, Lindsay\,38, and NGC419 the blue stars lying above the cluster MSTOs do not show
any obvious concentration toward the cluster centers and are fairly evenly distributed across our images. This supports
the interpretation that they are not BSS canditates belonging to these clusters, but instead part of the younger MS of 
the SMC field star population. We have used the Kolmogorov-Smirnov (KS) test to search for statistical differences 
between the cumulative projected radial distributions. The KS probabilities that BSS candidates and cluster sample 
stars are extracted from the same parent population are 17$\%$ (Lindsay\,1), 0$\%$ (Kron\,3), 0$\%$ (NGC\,339), 
4$\%$ (Lindsay\,38), and
25$\%$ (NGC\,419). For NGC\,416 and NGC\,419, the radial distributions are only shown for the HRC data.
Because BSSs are assumed to be located in the cluster center, and the field stars are already overdense in the 
BSS candidate region on the HRC CMD, the analysis of the HRC data instead of the entire sample is justified.
Even though we know that NGC\,121 contains BSSs and the radial distribution shows a concentration of the BSS 
candidates towards the cluster center, the KS-Test gives a probability of 16$\%$ that the two samples
are from the same distribution.  

Within the center region of NGC\,416 covered by the HRC data, we find an indication for centrally concentrated 
BSS candidates. The dashed line (BSS candidate sample) almost lies on top of the solid line (cluster sample). 
The cluster location is close to the SMC main body, hence some of the BSS candidates are likely field MS 
stars of the SMC. We will discuss the SMC field stellar population in greater 
detail in a separate paper. The KS probability for NGC\,416 is 85$\%$.

\begin{deluxetable}{cc}
\tablecolumns{2}
\tablewidth{0pc}
\tablecaption{BSS candidates}
\tablenote{Number of BSS candidates after removal of stars possibly affected by blends}
\tablehead{
\colhead{Cluster} & \colhead{Number of BSS candidates $^a$}} 
\startdata
Lindsay\,1  &  110 \\
Kron\,3     &  229 \\
NGC\,339    &  616 \\
NGC\,416    &  91  \\
Lindsay\,38 &  23  \\
NGC\,419    &  8   \\
\enddata
\label{tab:BSS}
\end{deluxetable}

\section{CLUSTER AGES}
\label{sec:age}

Age determinations of star clusters using isochrones depend crucially on the interstellar extinction, distance and 
metallicity of the cluster, as well as on the chosen isochrone models. We used spectroscopic metallicities  
\citep[][Kayser et al. 2008, in preparation]{daco98,Kayser06,Kayser07} in order to eliminate metallicity as a free 
parameter when fitting isochrone models. The distance
and the extinction were treated as free parameters. The mean SMC distance modulus is assumed to be $(m-M)_0$ = 
$18.88 \pm 0.1$~mag (60~kpc) \citep[e.g., ][]{Storm04}. Due to the large depth extension of the SMC and 
thus an expected wide spread in the cluster distances we adjusted the distance modulus for each star cluster 
to produce the best isochrone fits to our CMD data.

As mentioned before, we visually estimated the location of the cluster center on the image and selected all 
stars within an individually defined radius on the WFC data, except for NGC\,416 and NGC\,419 for which we 
used the HRC data. The center samples were used to fit the fiducial ridgelines for easier comparison to the 
isochrones. 

In order to fit ridgelines, we separated the cluster sample CMDs into three sections: the MS, the SGB and the RGB. 
On the MS we determined the mode of the color distribution in magnitude bins of 0.3~mag width. For the SGB, we 
performed a linear least squares fit of a polynomial of 5th order to a Hess diagram of this region in the CMD. 
Finally, the RGB was fit by a third-order polynomial of the mean color, again in magnitude bins with a size of 
0.3~mag each. The resulting ridgelines are shown as cyan lines in our CMDs and will be made available 
electronically (see Table~\ref{tab:ridgeline} for an illustrative excerpt). 

\begin{deluxetable}{cc}
\tablecolumns{4}
\tablewidth{0pc}
\tablecaption{Ridgeline of Lindsay\,1}
\tablenote{Table~\ref{tab:ridgeline} is presented in its entirety in the electronic edition of the Astronomical Journal. A
portion of the ridgeline of Lindsay\,1 is shown here. }
\tablehead{
\colhead{$m_{555}-m_{814}$} & \colhead{$m_{555}$} \\
\colhead{} & \colhead{} }
\startdata
1.3230 &  27.1000 \\
1.1720 &  26.5000 \\
1.0320 &  25.9000 \\
0.9090 &  25.3000 \\
0.7950 &  24.7000 \\
\enddata
\label{tab:ridgeline}
\end{deluxetable}

We fitted our $m_{555}$ vs $m_{555}$-$m_{814}$ CMDs with three different isochrone models: Padova isochrones 
(L.~Girardi 2006, private communication, Girardi et al. 2000)\footnote{$http://pleiadi.pd.astro.it/isoc\_photsys.02/isoc\_acs\_wfc/index.html$}, 
Teramo isochrones \citep{piet04} and Dartmouth isochrones \citep{dotter07}, all with scaled solar abundances
([$\alpha$/Fe]=0.0). The Padova isochrone grid has an age resolution of log(t)= 0.01, the Teramo isochrone grid 
of 0.1~Myr and the Dartmouth isochrone grid of 0.5~Gyr. The Padova isochrones model the AGB and its tip, which is 
$\sim$1~mag brighter than the tip of the RGB. The Teramo isochrones include the RGB, HB and the lower AGB. The Dartmouth 
isochrones terminate at the He flash, because they do not have the HB and the AGB sequence calculated. All sets 
of isochrones are available in the standard ACS color system.

We fitted a large number of isochrones using different combinations of reddening, age and distance. For each set of models,
we selected by trial and error the isochrone that best matched the observed data.


\subsection{Age of Lindsay\,1}
\label{sec:ages}

\begin{figure}
  \epsscale{1}
  \plotone{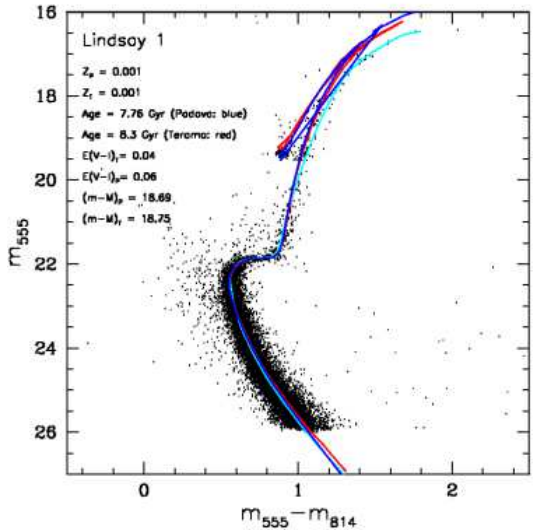}
  \caption{The CMD of Lindsay\,1 with the best-fitting isochrones of two different models: The blue solid line shows the 
  best-fitting Padova (L.Girardi, 2006, private communication, Girardi et al. 2000) isochrone that is closest to the 
  spectroscopically measured 
  metallicity of the cluster. The red solid line is the best-fitting Teramo \citep{piet04} isochrone approximating the 
  known metallicity. The cyan solid line is our fiducial ridgeline. The fitting parameters are listed in the plot legend.}
  \label{fig:l1_finaliso}
\end{figure}

\begin{figure}
  \epsscale{1}
  \plotone{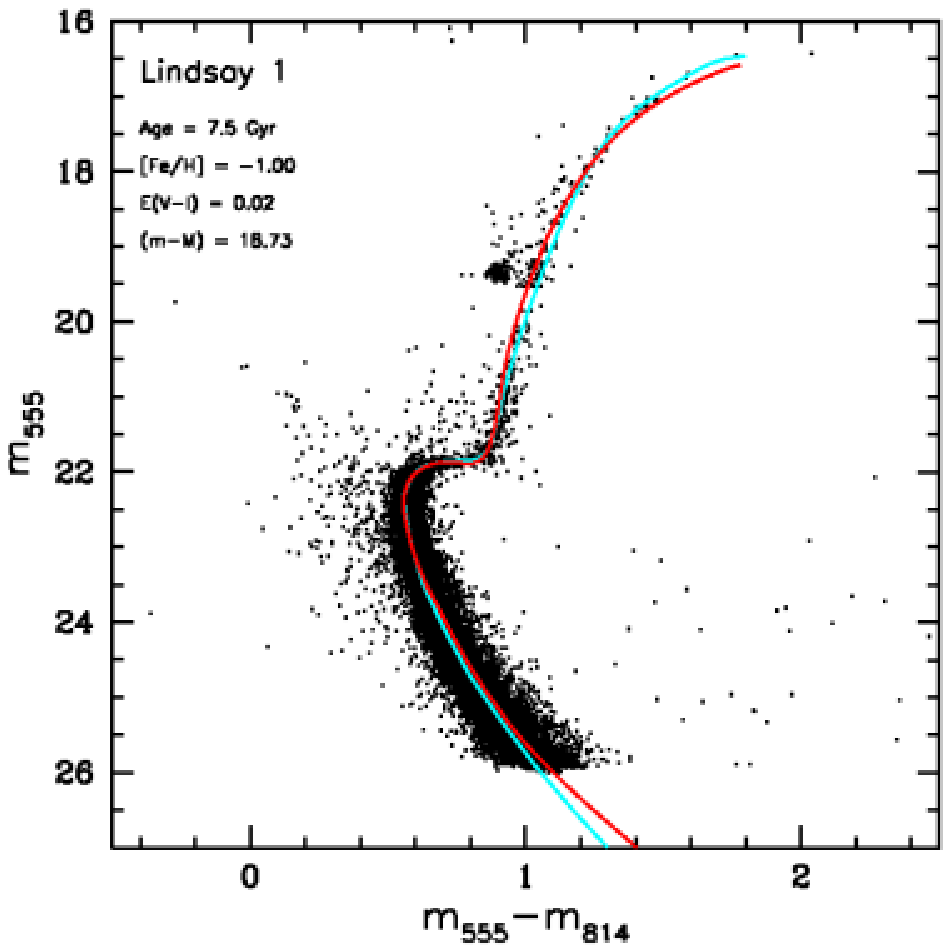}
  \caption{The Lindsay\,1 CMD with the best-fitting Dartmouth \citep{dotter07} isochrones overplotted in red. As before, the cyan
  line represents our fiducial for Lindsay\,1. The fit parameters are listed in the plot.}
  \label{fig:l1_dart}
\end{figure}

\begin{figure}
  \epsscale{1}
  \plotone{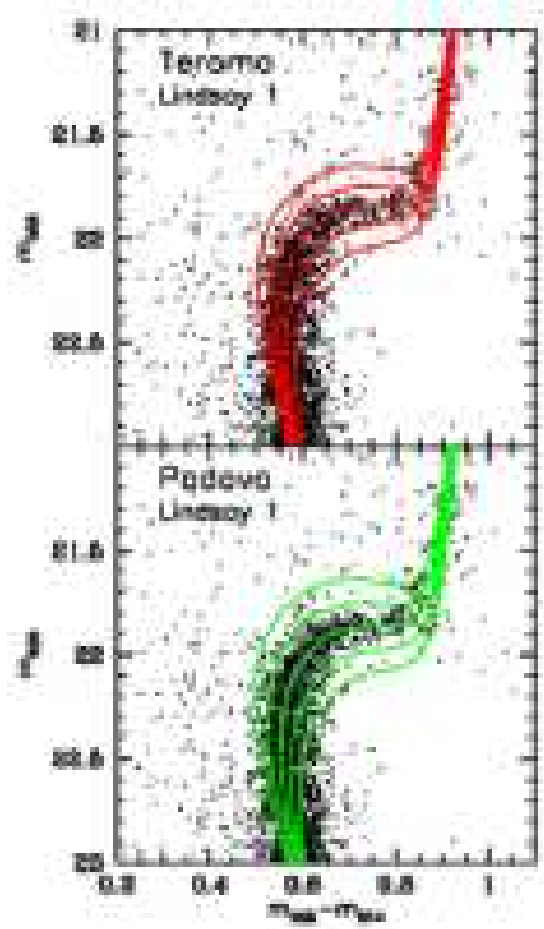}
 \caption{The CMD of Lindsay\,1 after zooming in on the region of the main-sequence turnoff, subgiant branch,
 and lower red giant branch. In the upper panel, we show Teramo isochrones as solid lines, covering an age range of 6.8, 7.5, 8.2, 
 9 and 10~Gyr. In the lower panel we show the same plot for Padova isochrones in the available age steps of 6.16, 6.92, 7.76, 8.7,
 and 9.77~Gyr. All other parameters are the same as in Figs.~\ref{fig:l1_finaliso} and~\ref{fig:l1_dart}.}
 \label{fig:l1_isocheck}
\end{figure}

We adopted the metallicity of [Fe/H] = -1.14$\pm$0.10 from \citet{daco98} in the scale of \citet{zinn84} (ZW84).
This metallicity corresponds most closely to Z = 0.001 in both the Padova and the Teramo models. Our best-fit
age using Padova isochrones is 7.76~Gyr with $(m-M)_0$ = 18.69~mag and $E_{V-I}$ = 0.06. The best fitting Teramo 
isochrone yields an age of t = 8.3~Gyr, $(m-M)_0$ = 18.75~mag, and $E_{V-I}$ = 0.04 (see Fig.~\ref{fig:l1_finaliso}). 

On the MS, the Padova isochrone is slightly offset to the red by about $\sim$0.1~mag in color, while the Teramo 
isochrone is slightly offset by about $\sim$0.02~mag in color. Between $22 \lesssim m_{555} \lesssim $ 23~mag, 
both the Padova isochrone and the Teramo isochrone 
are shifted by $\sim$0.05~mag to the blue with respect to the main ridgeline. Both isochrones provide an excellent 
approximation to the SGB and to the lower RGB up to about 1~mag below the red clump. At brighter magnitudes, the two 
isochrones deviate increasingly to the blue of the observed upper RGB. Here the Padova isochrones show the strongest 
difference, deviating by approximately 0.24~mag in color from the observed tip of the RGB. The Teramo isochrone is too 
blue by about 0.13~mag at the magnitude of the tip of the RGB.

\begin{figure}
  \epsscale{1}
  \plotone{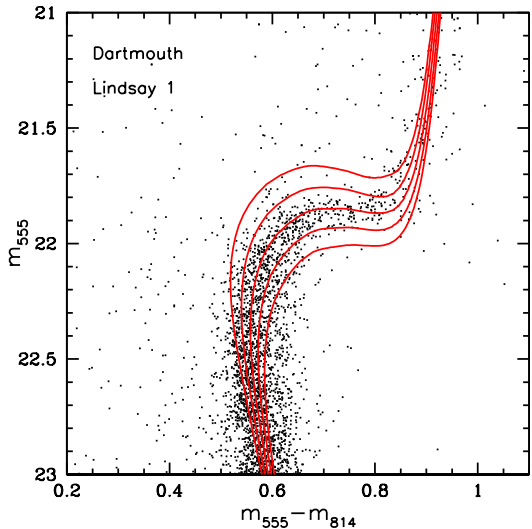}
 \caption{Same as Fig.~\ref{fig:l1_isocheck}, but for the Dartmouth isochrones with ages of 6.5, 7, 
 7.5, 8 and 8.5~Gyr.}
 \label{fig:l1_darth_isocheck}
\end{figure}

The best fit using the Dartmouth isochrones is obtained with 7.5~Gyr, $(m-M)_0$ = 18.73~mag and $E_{V-I}$ = 0.02 for an 
isochrone corresponding to a metallicity [Fe/H] = -1.00 (Fig.~\ref{fig:l1_dart}). As for the Padova and the Teramo isochrones,
on the MS the isochrone is $\sim$0.02~mag offset to the red. The SGB is fitted to a very well and the deviation 
on the RGB is much smaller than for the other two isochrone models. On the upper RGB, the isochrone is increasingly 
offset to the red relative to the fiducial ridgeline. The derived reddening by the Dartmouth isochrones agrees best with 
the extinction $A_V$ = $0.1$ from the \citet{Schlegel98} maps ($E_{V-I}$ = $0.024$~mag). The reddening law of 
\citet{Dean78} is assumed. 

In the Figures~\ref{fig:l1_isocheck} and~\ref{fig:l1_darth_isocheck} we show a range of isochrones for the three sets of stellar
evolution models in order to illustrate the age uncertainty in a given model. The best fit isochrone is always displayed 
along with two younger and two older isochrones. 
Due to the high quality of the fit of the central isochrone in the region of the upper MS, SGB, and lower
RGB in all models and the larger deviations of the adjacent isochrones, we estimate that the resultant age uncertainty is 
of the order of approximately $\pm$ 0.7~Gyr for the Teramo and Padova isochrones and $\pm$ 0.5~Gyr for the Dartmouth 
isochrones.


\subsection{Age of Kron\,3}

\begin{figure}
  \epsscale{1}
  \plotone{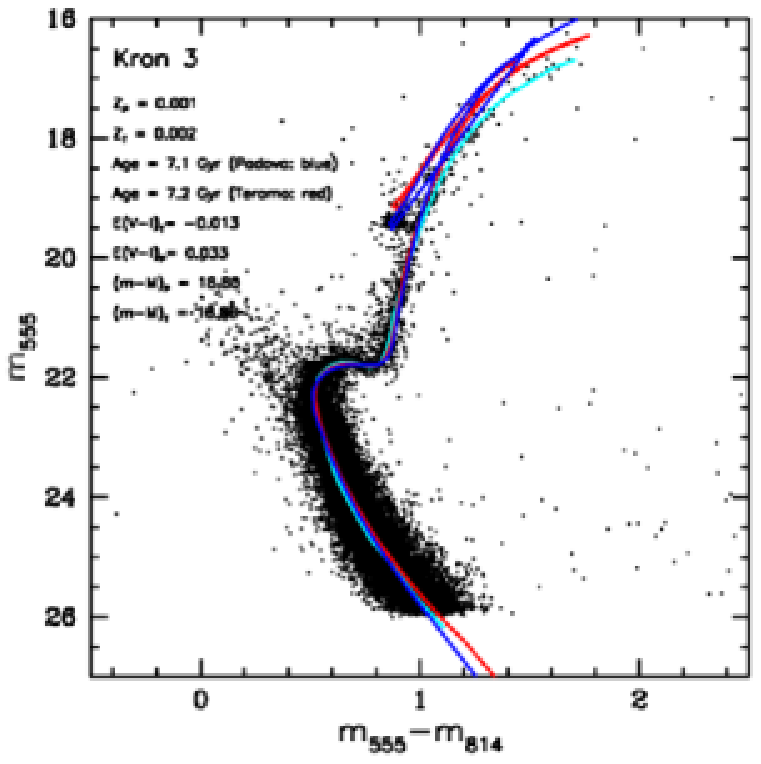}
  \caption{The CMD of Kron\,3 with the best-fitting isochrones of two different models: The blue solid line shows the 
  best-fitting Padova isochrone that is closest to the spectroscopically measured metallicity of the cluster. The
  red solid line is the best-fitting Teramo isochrone approximating the known metallicity. The cyan solid line is our
  fiducial ridgeline. The fit parameters are listed in the plot legend.}
  \label{fig:k3_finaliso}
\end{figure}

\begin{figure}
  \epsscale{1}
  \plotone{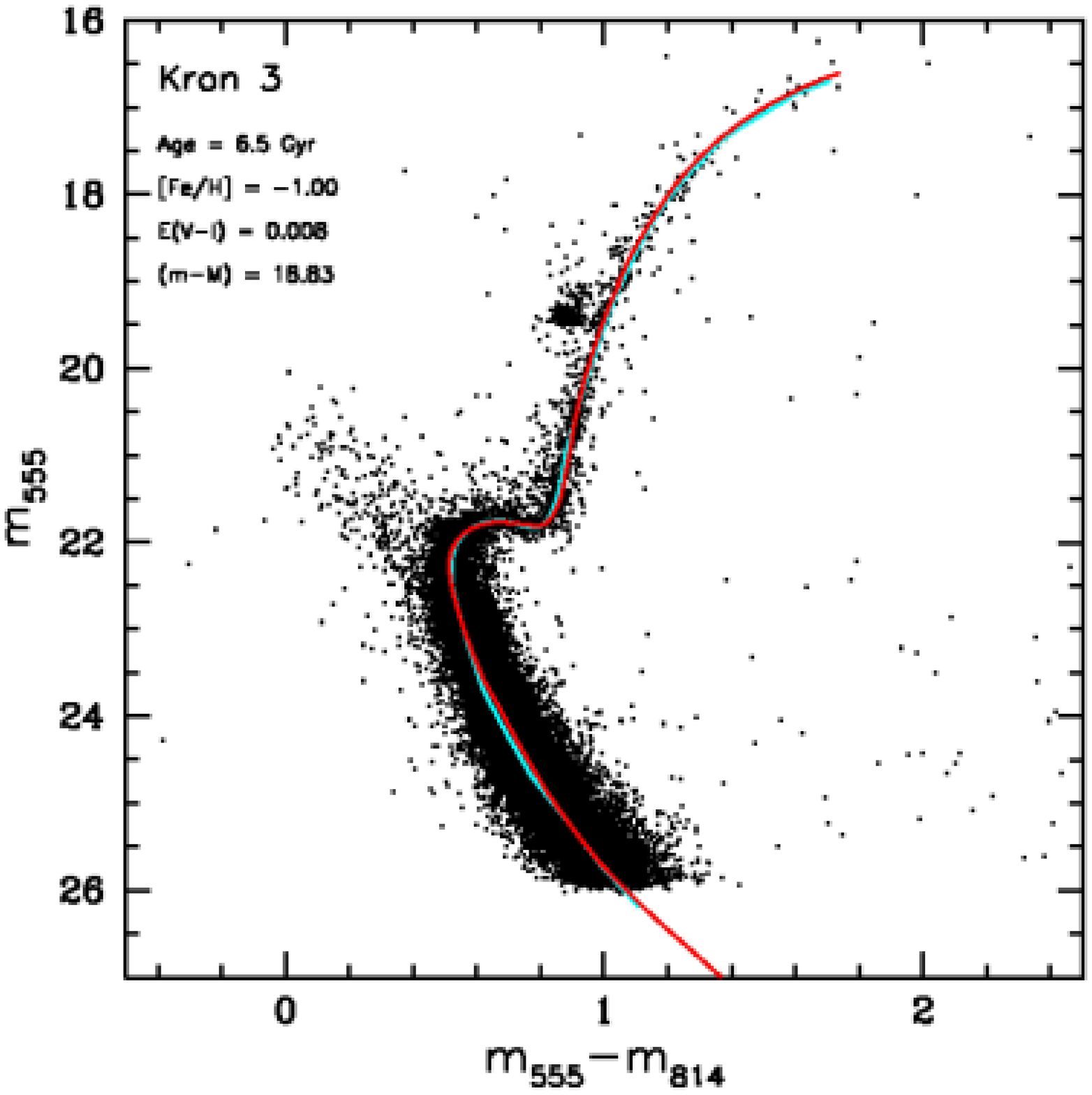}
  \caption{The Kron\,3 CMD with the best-fitting Dartmouth isochrones overplotted in red. As before, the cyan
  line represents our fiducial for Kron\,3. The fit parameters are listed in the plot.}
  \label{fig:k3_dart}
\end{figure}

\begin{figure}
  \epsscale{1}  
  \plotone{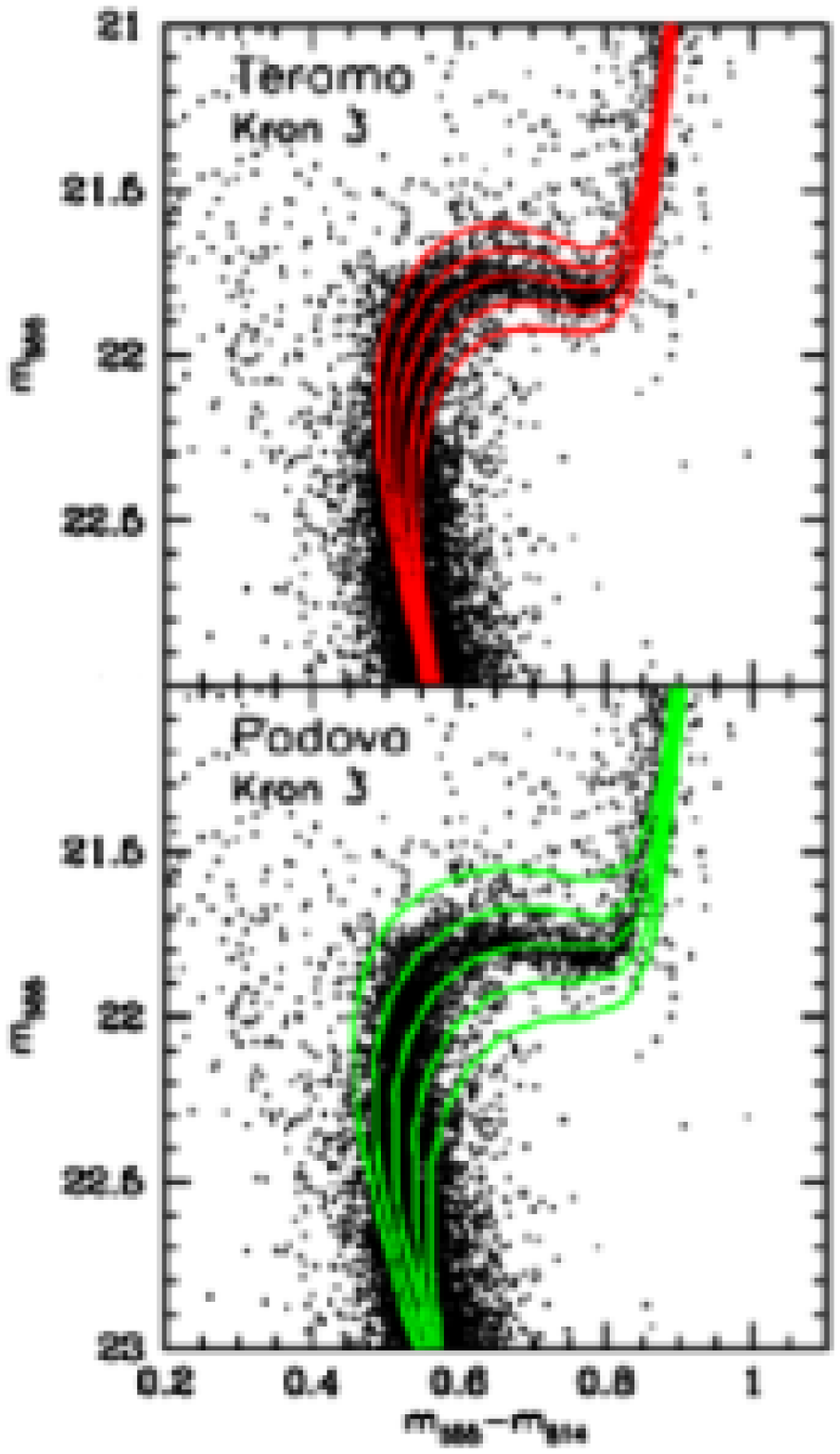}
 \caption{The CMD of Kron\,3 after zooming in on the region of the main-sequence turnoff, subgiant branch,
 and lower red giant branch. In the upper panel, we show Teramo isochrones as solid lines, covering an age range of 
 5.2, 6.1, 7, 7.8, and 8.4~Gyr. In the lower panel we show the same plot for Padova isochrones in the available age steps of 
 5.6, 6.3, 7.1, 7.9, and 8.9~Gyr. All other parameters are the same as in Figs.~\ref{fig:k3_finaliso} and~\ref{fig:k3_dart}.}
 \label{fig:k3_isocheck}
\end{figure}

We adopted the spectroscopic metallicity derived by \citet{daco98} of [Fe/H] = $-1.08 \pm 0.12$. This metallicity corresponds 
most closely to Z=0.001 in the Padova models, Z=0.002 in the Teramo models and [Fe/H] = -1.00 in the Dartmouth models.
The best-fit age using the Padova isochrones was found to be 7.1~Gyr, $(m-M)_0$ = 18.80~mag and $E_{V-I}$ = 0.033. The best 
fit for the Teramo isochrones is obtained with t = 7.2 Gyr, $(m-M)_0$ = 18.80~mag and $E_{V-I}$ = -0.013 
(Fig.~\ref{fig:k3_finaliso}). No reasonable fit with a zero or positive reddening value was obtained.

On the MS, both isochrones are slightly offset to the red by about $\sim$0.02 (Teramo) and $\sim$0.01 (Padova) in color.  
The Teramo isochrone traces the SGB and the lower RGB to a high accuracy. Only at the faint end of the SGB the isochrone 
is about $\sim$0.02~mag brighter than our fiducial ridgeline. The isochrone deviates increasingly to the blue of the observed 
upper RGB, but fits the color of the RGB tip very well with a small deviation of about 0.02~mag. The base of the red clump, 
however, is $\sim$0.4~mag too bright.
The Padova isochrone lies $\sim$0.1~mag below our observed SGB and is $\sim$0.02~mag offset in color on the lower RGB.
The isochrone continues further to the blue on the upper RGB as seen for the Teramo isochrone. The Padova isochrone shows the 
strongest difference at the RGB tip, deviating by approximately 0.22~mag in color from the observed tip of the RGB. The base 
of the red clump is fitted very well. 

\begin{figure}
  \epsscale{1}
  \plotone{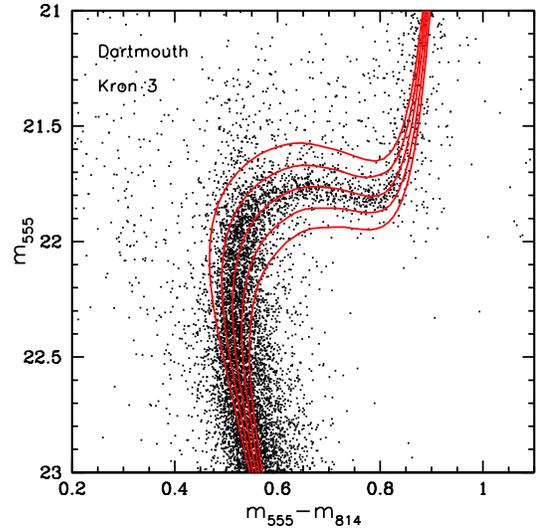}
 \caption{Same as Fig.~\ref{fig:k3_isocheck}, but for the Dartmouth isochrones with ages of 5.5, 6, 
 6.5, 7, and 7.5~Gyr.}
 \label{fig:k3_dart_isocheck}
\end{figure}

The Dartmouth isochrone model provided by \citet{dotter07} yields the best fit to the CMD (Fig.~\ref{fig:k3_dart}). 
The best-fit isochrone has the parameters t = 6.5~Gyr, $(m-M)_0$ = 18.83 and $E_{V-I}$ = 0.008. All the major 
features of the CMD are very well reproduced, including the upper RGB where the isochrone is offset slightly to the 
blue relative to the fiducial ridgeline. This offset is no more than 0.02~mag on average along the entire upper RGB. 
Even the slope of the RGB is very well reproduced along its entire extent. The derived reddening for Padova isochrones 
agrees best with the extinction $A_V$ = $0.09$ from the \citet{Schlegel98} maps ($E_{V-I}$ = $0.024$~mag).

In Figures~\ref{fig:k3_isocheck} and~\ref{fig:k3_dart_isocheck} we display our ''best'' isochrone along with two older 
and two younger ones for each model. Here the deviations of these isochrone models from the observed upper MS and SGB 
are very clearly visible. 
We estimate the resultant 
age uncertainty in order of approximately $\pm$ 0.5~Gyr for the Teramo and Dartmouth isochrones and $\pm$ 0.7~Gyr for the Padova 
isochrones.


\subsection{Age of NGC\,339}

\begin{figure}
  \epsscale{1}
  \plotone{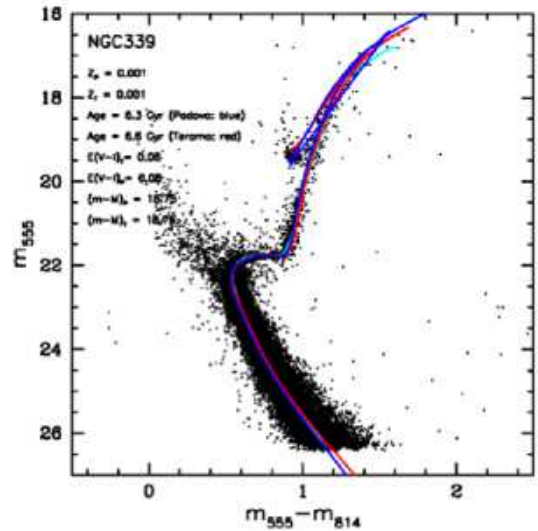}
  \caption{The CMD of NGC\,339 with the best-fitting isochrones of two different models: The blue solid line shows the 
  best-fitting Padova isochrone that is closest to the spectroscopically measured metallicity of the cluster. The
  red solid line is the best-fitting Teramo isochrone approximating the known metallicity. The cyan solid line is our
  fiducial ridgeline. The fit parameters are listed in the plot legend.}
  \label{fig:ngc339_finaliso}
\end{figure}

\begin{figure}
  \epsscale{1}
  \plotone{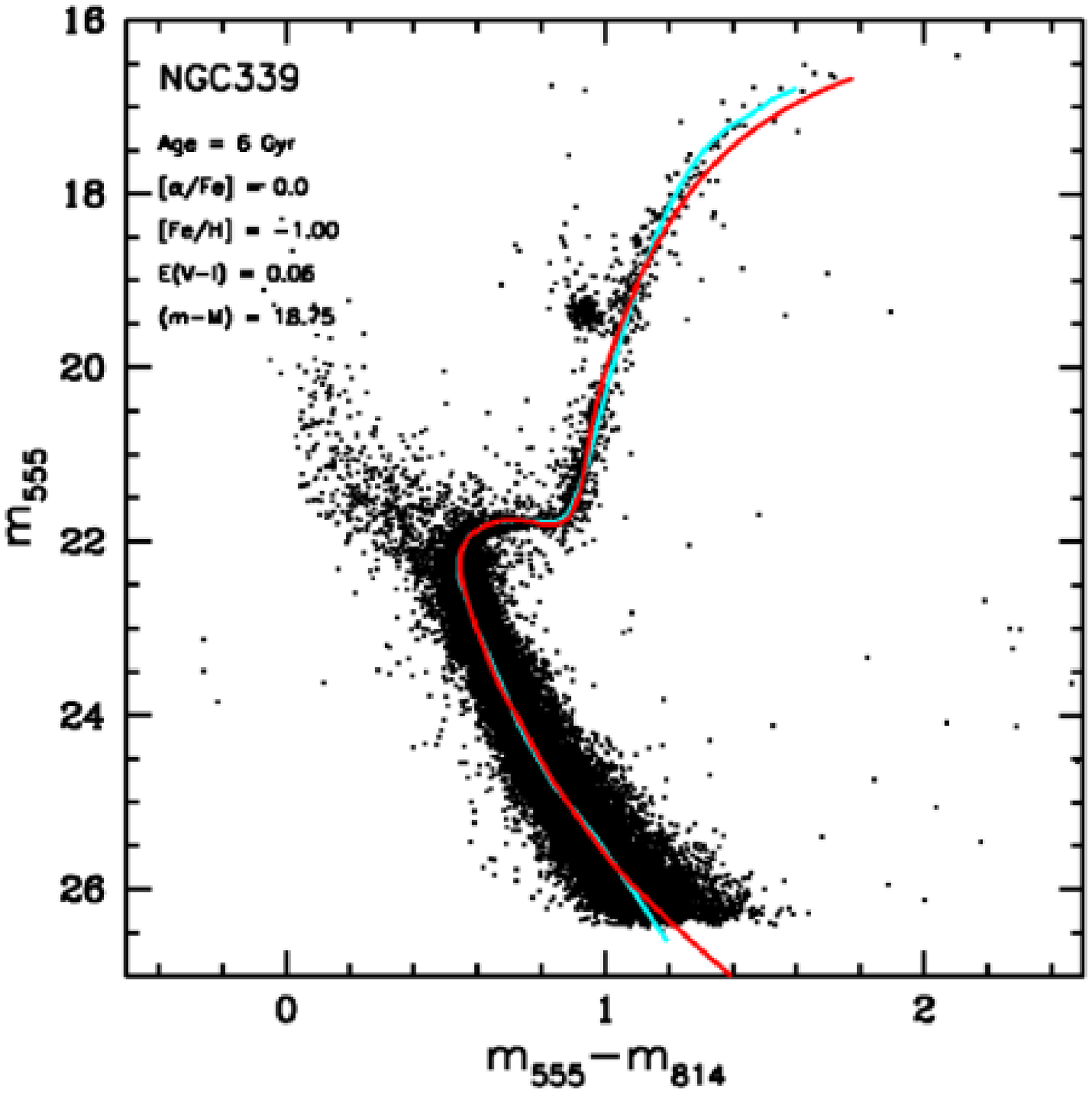}
  \caption{The NGC\,339 CMD with the best-fitting Dartmouth isochrones overplotted in red. As before, the cyan
  line represents our fiducial for NGC\,339. The fit parameters are listed in the plot.}
  \label{fig:ngc339_dart}
\end{figure}

\begin{figure}
  \epsscale{1}  
  \plotone{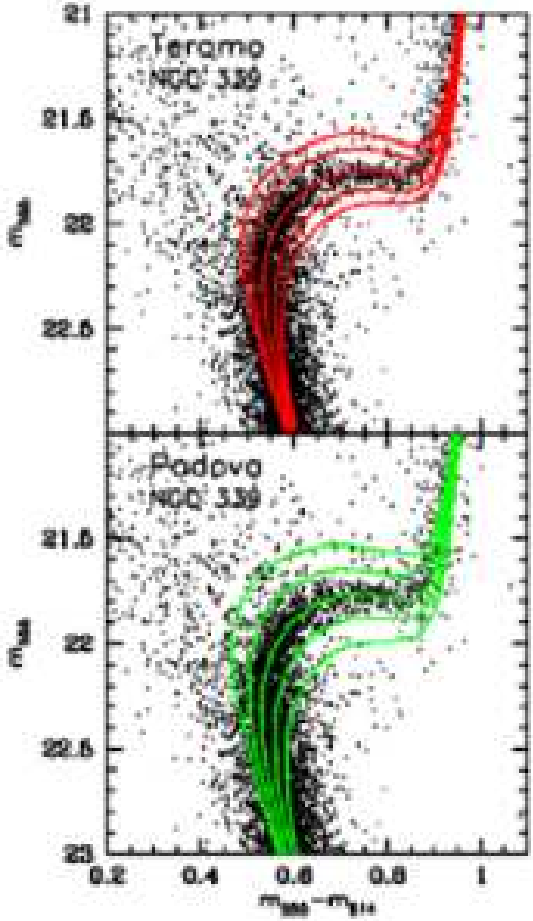}
 \caption{The CMD of NGC\,339 after zooming in on the region of the main-sequence turnoff, subgiant branch,
 and lower red giant branch. In the upper panel, we show Teramo isochrones as solid lines, covering an age range of 5.6, 6, 
 6.6, 7.2, and 7.7~Gyr. In the lower panel we show the same plot for Padova isochrones in the available age steps of  5, 5.6, 
 6.3, 7, and 7.9~Gyr. All other parameters are the same as in Figs.~\ref{fig:ngc339_finaliso} and~\ref{fig:ngc339_dart}.}
 \label{fig:ngc339_isocheck}
\end{figure}

We adopted a spectroscopic metallicity of [Fe/H] = $-1.12 \pm 0.10$ from \citet{daco98}, which made us use 
isochrones with Z = 0.001 for both the Teramo and the Padova models. The best age fit using the Teramo isochrone 
is found with the parameters t = 6.6~Gyr, $(m-M)_0$ = 18.75~mag and $E_{V-I}$ = 0.08. Our best fitting Padova 
isochrone yields an age of t = 6.3~Gyr with $(m-M)_0$ = 18.75~mag and $E_{V-I}$ = 0.08. 

Both isochrones provide an excellent approximation to all features of the CMD (Fig.~\ref{fig:ngc339_finaliso}), 
the MS, the SGB and even the upper RGB, where the isochrones are only slightly offset to the blue relative to 
the fiducial ridgeline. These offsets are no more than 0.01~mag (Teramo) and 0.03~mag (Padova) in 
color on average along the entire upper RGB. Also the slope of the RGB is very well reproduced along its 
entire extent. At the RGB tip, the Padova isochrone deviates by approximately 0.15~mag in color to the red
from the observed RGB tip, while the Teramo isochrone match the observed color and luminosity of the tip
better, deviating $\sim$ 0.10~mag in color to the red. The Teramo isochrone 
shows a magnitude for the base of the red clump that is about 0.2 brighter than the observed one. The Padova 
isochrone indicates a magnitude for the base of the red clump that is 0.2~mag too faint.

Figure~\ref{fig:ngc339_dart} shows the best-fit isochrone for the Dartmouth model. The best fit is obtained with 
an age of t = 6~Gyr and the parameters $(m-M)_0$ = 18.75~mag and $E_{V-I}$ = 0.06. The Dartmouth isochrone traces 
the ridgeline on the MS, the SGB and the lower RGB to a very well. On the lower RGB, the isochrone is 
offset to the blue by $\sim$0.03~mag, while on the upper RGB it is $\sim$0.05~mag too red. The derived reddening 
for Dartmouth isochrones agrees best with the extinction $A_V$ = $0.15$ from the \citet{Schlegel98} maps 
($E_{V-I}$ = $0.04$~mag).

\begin{figure}
  \epsscale{1}
  \plotone{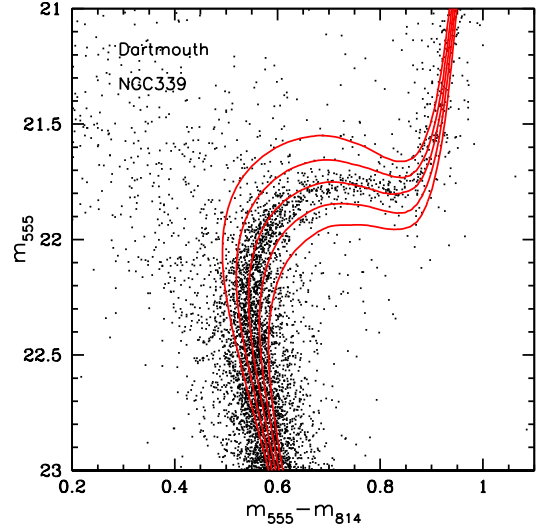}
 \caption{Same as Fig.~\ref{fig:ngc339_isocheck}, but for the Dartmouth isochrones with ages of 5, 5.5, 
 6, 6.5, and 7~Gyr.}
 \label{fig:ngc339_dart_isocheck}
\end{figure}

In Figures~\ref{fig:ngc339_isocheck} and~\ref{fig:ngc339_dart_isocheck} we show a range of isochrones for each 
model to illustrate the age uncertainty. We always display our ''best'' fit isochrone along with two older and 
two younger ones. Considering the excellent fit of all age-sensitive features of the CMD by all three stellar 
evolution models, we estimate the resultant age uncertainty to be approximately $\pm$0.5~Gyr for all three models.


\subsection{Age of NGC\,416}

\begin{figure}
  \epsscale{1}
  \plotone{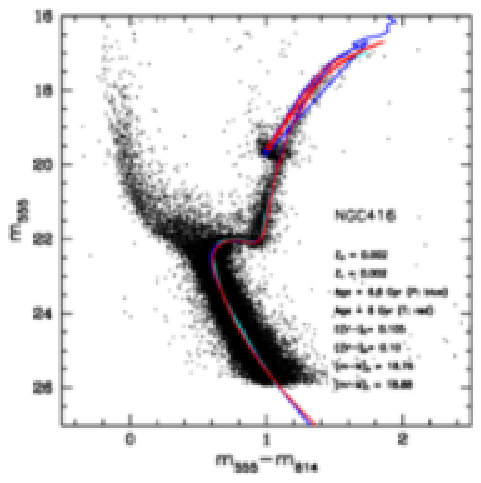}
  \caption{The CMD of NGC\,416 with the best-fitting isochrones of two different models: The blue solid line shows the best-fitting
  Padova (L.Girardi, 2006, in preparation) isochrone that is closest to the spectroscopically measured metallicity of the cluster. The
  red solid line is the best-fitting Teramo \citep{piet04} isochrone approximating the known metallicity. The cyan solid line is our
  fiducial ridgeline. The fit parameters are listed in the plot legend.}
  \label{fig:ngc416_finaliso}
\end{figure}

\begin{figure}
  \epsscale{1}
  \plotone{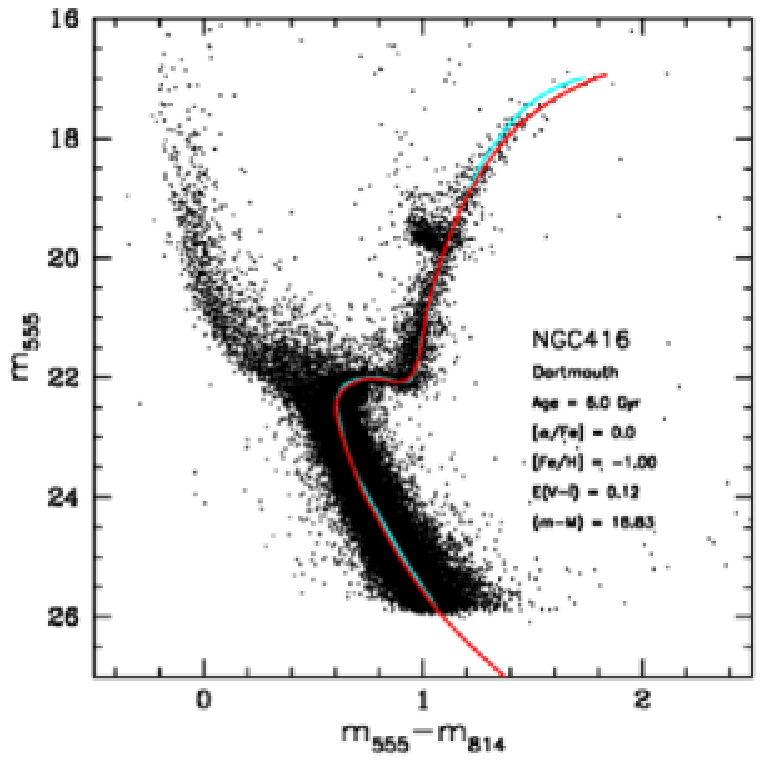}
  \caption{The NGC\,416 CMD with the best-fitting Dartmouth \citep{dotter07} isochrones overplotted in red. As before, the cyan
  line represents our fiducial for NGC\,416. The fit parameters are listed in the plot.}
  \label{fig:ngc416_dart}
\end{figure}

\begin{figure}
  \epsscale{1}  
  \plotone{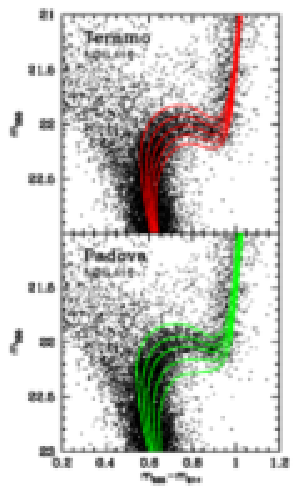}
 \caption{The CMD of NGC\,416 after zooming in on the region of the main-sequence turnoff, subgiant branch,
 and lower red giant branch. In the upper panel, we show Teramo isochrones as solid lines, covering an age range of  5, 5.5, 6, 6.5, 
 and 7 ~Gyr. In the lower panel we show the same plot for Padova isochrones in the available age steps of 5.4, 6.0, 6.6, 7.4, 
 and 8.3~Gyr. All other parameters are the same as in Figs.~\ref{fig:ngc416_finaliso} and~\ref{fig:ngc416_dart}.}
 \label{fig:ngc416_isocheck}
\end{figure}

The latest spectroscopic metallicity measurement was performed by Kayser et al. (2008, in preparation) and yields a 
metallicity $[Fe/H]_{CG97}$ = $-0.87 \pm 0.06$ in the scale of \citet{cargra97} (CG97) . We transform this metallicity 
to the ZW84 scale using the transformation given by \citet{cargra97}, and obtain a metallicity of $[Fe/H]_{ZW84}$ = 
$-1.00$. This metallicity corresponds most closely to Z = 0.002 for both the Teramo and the Padova models. In 
Figure~\ref{fig:ngc416_finaliso}, the CMD with the overplotted Teramo and Padova isochrones is shown. The best fit 
for the Teramo isochrones is obtained with t = 6.0~Gyr, $(m-M)_0$ = 18.88~mag and $E_{V-I}$ = 0.105, while the Padova 
isochrones provide t = 6.6~Gyr, $(m-M)_0$ = 18.76~mag and $E_{V-I}$ = 0.10. The relatively high reddening value is 
due to the location of the cluster close to the SMC main body (see Fig.~\ref{fig:clusters}).

On the MS, both the Teramo isochrone and the Padova isochrone are offset to the blue by about $\sim$0.01~mag.
While the Teramo isochrone fits the MSTO rather nicely, the Padova isochrone is $\sim$0.05~mag too blue. At the blue 
end of the SGB, both isochrones are slightly too faint. The Teramo isochrone fits the red part of the SGB and the 
entire RGB almost perfectly up to about 0.5 magnitudes below the RGB tip (see Fig.~\ref{fig:ngc416_finaliso}).
The Padova isochrone fits the red end of the SGB almost perfectly, but deviates increasingly to the blue on 
the RGB with respect to our fiducial ridgeline. The isochrone shows a magnitude for the base of the red clump 
that is about 0.1~mag brighter than the observed one. The Teramo isochrone indicates a magnitude for the base 
of the red clump that is 0.25~mag too bright.

\begin{figure}
  \epsscale{1}
  \plotone{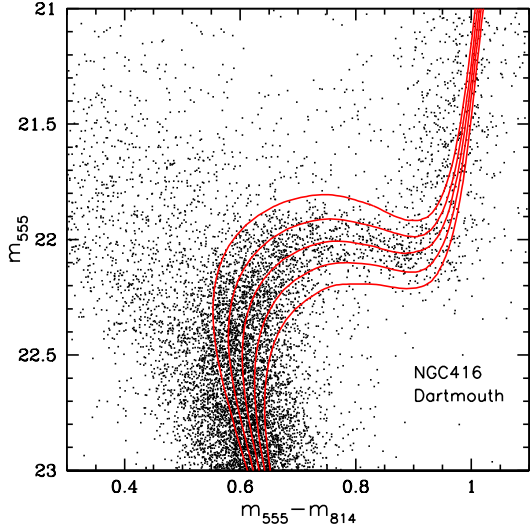}
 \caption{Same as Fig.~\ref{fig:ngc416_isocheck}, but for the Dartmouth isochrones with ages of 5, 
 5.5, 6, 6.5, and 7~Gyr.}
 \label{fig:ngc416_dart_isocheck}
\end{figure}

In Figure~\ref{fig:ngc416_dart} the best fitting isochrone for the Dartmouth models \citep{dotter07} is 
displayed with t = 6~Gyr, $(m-M)_0$ = 18.83~mag and $E_{V-I}$ = 0.12. We adopted the isochrone set with a 
metallicity of [Fe/H] = $-1.0$. All age-sensitive features of the CMD are well reproduced. On the upper RGB, 
the isochrone is increasingly offset to the blue relative to the fiducial ridgeline with 0.03~mag in color 
being the strongest difference. Our derived reddenings agree with the extinction $A_V$ = $0.32$ from the 
\citet{Schlegel98} maps ($E_{V-I}$ = $0.1$~mag). 

In Figures~\ref{fig:ngc416_isocheck} and~\ref{fig:ngc416_dart_isocheck} a range of five isochrones is displayed 
for each isochrone model. Even though the approximation of all features that are important for the age 
determination important features is close, we estimate the age uncertainty to be $\sim$0.8~Gyr. It is
possible that the broad width of the RGB suggests a spread in metallicity and therefore in age, 
which we have to take into account. The main reason for the broadening of the RGB, however, are SMC field stars.


\subsection{Age of Lindsay\,38}

\begin{figure}
  \epsscale{1}
  \plotone{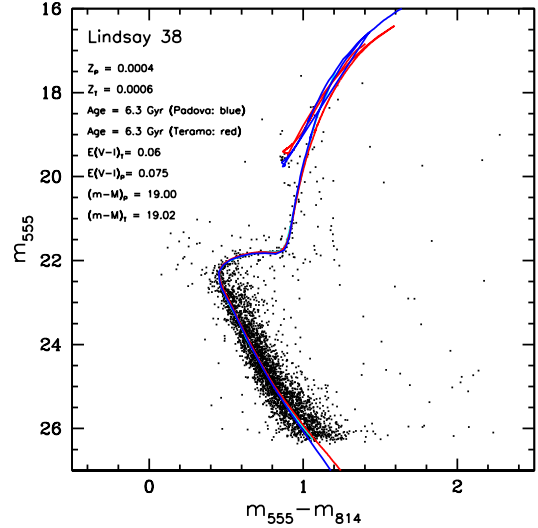}
  \caption{The CMD of Lindsay\,38 with the best-fitting isochrones of two different models: The blue solid line shows the 
  best-fitting Padova isochrone that is closest to the spectroscopically measured metallicity of the cluster. The
  red solid line is the best-fitting Teramo isochrone approximating the known metallicity. The cyan solid 
  line is our fiducial ridgeline. The fit parameters are listed in the plot legend.}
  \label{fig:l38_finaliso}
\end{figure}

\begin{figure}
  \epsscale{1}
  \plotone{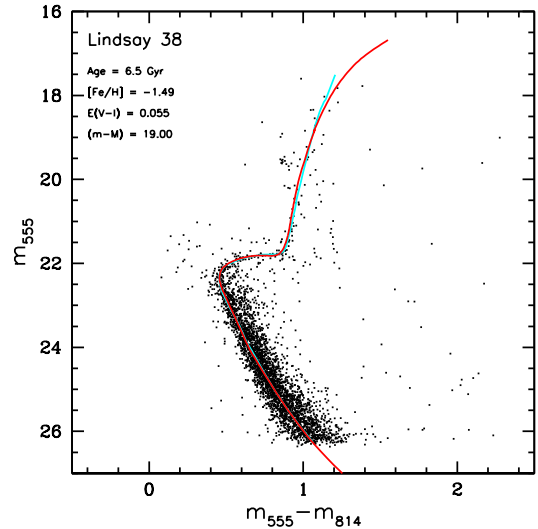}
  \caption{The Lindsay\,38 CMD with the best-fitting Dartmouth isochrones overplotted in red. As before, the cyan
  line represents our fiducial for Lindsay\,38. The fit parameters are listed in the plot.}
  \label{fig:l38_dart}
\end{figure}

\begin{figure} 
  \epsscale{1}  
  \plotone{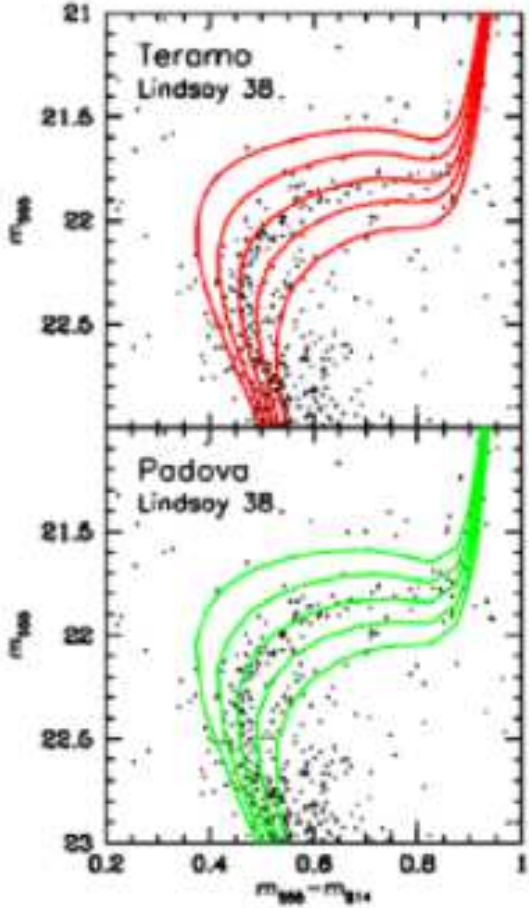}
 \caption{The CMD of Lindsay\,38 after zooming in on the region of the main-sequence turnoff, subgiant branch,
 and lower red giant branch. In the upper panel, we show Teramo isochrones as solid lines, covering an age range of 5, 5.6, 
 6.3, 7, and 8~Gyr. In the lower panel we  show the same plot for Padova isochrones in the available age steps of 5, 5.6, 
 6.3, 7.1, and 7.9~Gyr. All other parameters are the same as in Figs.~\ref{fig:l38_finaliso} and~\ref{fig:l38_dart}.}
 \label{fig:l38_isocheck}
\end{figure}

The latest measured spectroscopic metallicity provided by \citet[][2008 in prep.]{Kayser06,Kayser07}
is $[Fe/H]_{CG97}$ = $-1.35 \pm 0.10$, which we transformed to $[Fe/H]_{ZW84}$ = $-1.59$ \citep{cargra97}. 
This metallicity is in excellent agreement with the photometric metallicity found by \citet{pia01}. We used
isochrones of Z = 0.0004 in the Padova models, Z = 0.0006 in the Teramo models and [Fe/H] = -1.49 in the Dartmouth 
models. The best-fit age using the Teramo isochrone is t = 6.3~Gyr with  $(m-M)_0$ = 19.02~mag and $E_{V-I}$ = 0.06. 
The best fitting Padova isochrone yields an age of t = 6.3~Gyr, $(m-M)_0$ = 19.00 and $E_{V-I}$ = 0.075 
(Fig.~\ref{fig:l38_finaliso}). Surprisingly, Lindsay\,38 is the only cluster for which we found a high
quality fit using $\alpha$-enhanced Dartmouth isochrones ([$\alpha$/Fe] = +0.20), which yield an age of
6~Gyr using the same fitting parameters.

All features of the CMD are traced very well by both the Teramo and the Padova isochrones. At the faint end of 
the MS, the Teramo isochrone continues further to the red than the Padova isochrone and our derived ridgeline; 
however, this only becomes more apparent at magnitudes of $m_{555}$ = 25.5~mag and below. The MS, the SGB and 
the lower RGB are very well reproduced. The upper part of the RGB is too sparse for the fit of a ridgeline.
Therefore, a statement about the quality of the theoretical fiducials cannot be made. The base of the red clump 
for the Padova isochrone is about 0.2~mag too faint, while for the Teramo isochrone it is about 0.4~mag too bright.

In Figure~\ref{fig:l38_dart} the best-fit age provided by the Dartmouth isochrone \citep{dotter07} is shown with 
t = 6.5~Gyr, $(m-M)_0$ = 18.94~mag and $E_{V-I}$ = 0.05 (red line in Fig.~\ref{fig:l38_dart}). The MS and the SGB 
are very well reproduced. The lower RGB is slightly offset by about 0.02 on average along the entire RGB. The 
isochrone deviates increasingly to the red starting at 1.5~mag above the red clump. The reddening value found 
using isochrones is too high compared with the extinction taken from the \citet{Schlegel98} maps ($A_V$ = $0.05$, 
$E_{V-I}$ = $0.016$~mag). 

\begin{figure}
  \epsscale{1}
  \plotone{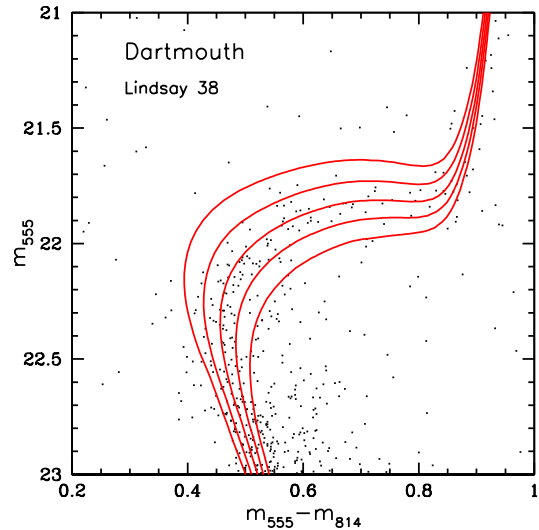}
 \caption{Same as Fig.~\ref{fig:l38_isocheck}, but for the Dartmouth isochrones with ages of 5.5, 6, 
 6.5, 7, and 7.5~Gyr.}
 \label{fig:l38_dart_isocheck}
\end{figure}

In Figures~\ref{fig:l38_isocheck} and~\ref{fig:l38_dart_isocheck} we show a selection of isochrones to estimate the age 
uncertainty. For each isochrone model two older and two younger isochrones are shown along with the ''best'' one. 
We estimate the age uncertainty to be $\pm$ 0.5~Gyr for all of the three isochrone models.

\subsection{Age of NGC\,419}
\label{sec:agen419}

\begin{figure}
  \epsscale{1}
  \plotone{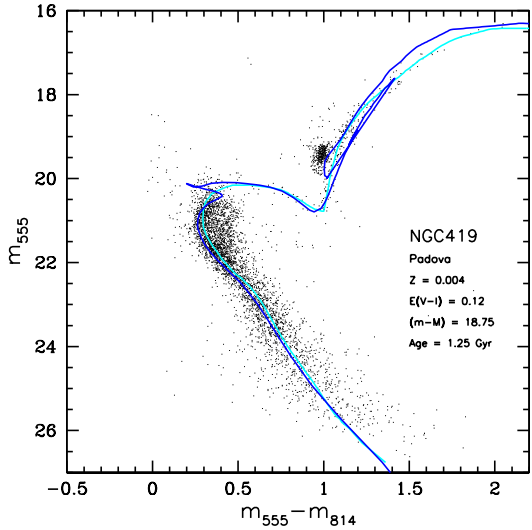}
 \caption{The CMD of NGC\,419 with the best-fitting isochrone of the Padova model: the blue solid line shows the best-fitting
  Padova isochrone that is closest to the spectroscopically measured metallicity of the 
  cluster. The distinction of the cluster from the field population is difficult due to the multiple turnoffs and the sparse SGB.}
 \label{fig:n419_padova}
\end{figure}

\begin{figure}
  \epsscale{1}
  \plotone{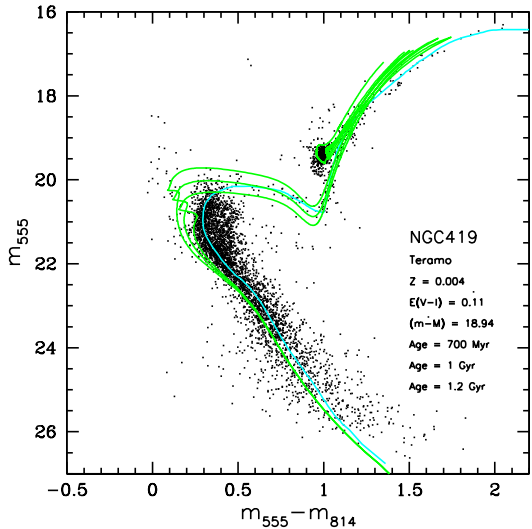}
 \caption{The CMD of NGC\,419 with the Teramo isochrones overplotted in green. As before, the cyan line represents
 our fiducial for NGC\,419. We show isochrones in the age range of 0.8, 1, 1.2~Gyr. The fit parameters are listed in the plot. }
 \label{fig:n419_teramo}
\end{figure}

The latest spectroscopic 
metallicity measurement was performed by Kayser et al. (2008, in preparation) and yields a metallicity 
$[Fe/H]_{CG97}$ = $-0.71 \pm 0.12$ in the CG97 scale. We transform this metallicity to the ZW84 scale using 
the transformation given by \citet{cargra97}, and obtain a metallicity of $[Fe/H]_{ZW84}$ = $-0.67$. This 
metallicity corresponds most closely to Z = 0.004 (Padova and Teramo) and $[Fe/H]$ = $-0.5$ (Dartmouth). 
The best fit for the Padova isochrones provides t = 1.25~Gyr with $(m-M)_0$ = 18.75~mag and $E_{V-I}$ = 0.12 
(see Fig.~\ref{fig:n419_padova}). The best-fit age using the Dartmouth isochrones is t = 1.5~Gyr with $(m-M)_0$ = 
18.60~mag and $E_{V-I}$ = 0.07. The Teramo model has problems with fitting all cluster features simultaneously 
(see Fig.~\ref{fig:n419_teramo}). The best-fitting age is t = 1~Gyr with $(m-M)_0$ = 18.94~mag and 
$E_{V-I}$ = 0.11. Note that it is not obvious where the exact location of NGC419's MSTO is.

As for NGC\,416, we find a relatively high reddening parameter due to the cluster location close to the SMC main 
body, where also a lot of crowding is expected. The Padova isochrone fits the MS and the SGB very well, while 
the lower RGB is offset by $\sim$0.05~mag in color, and its slope is not fitted at all. The isochrone does not 
fit the red clump, lies $\sim$0.02~mag to the red and is $\sim$0.6~mag too faint.

\begin{figure}
 \epsscale{1}
 \plotone{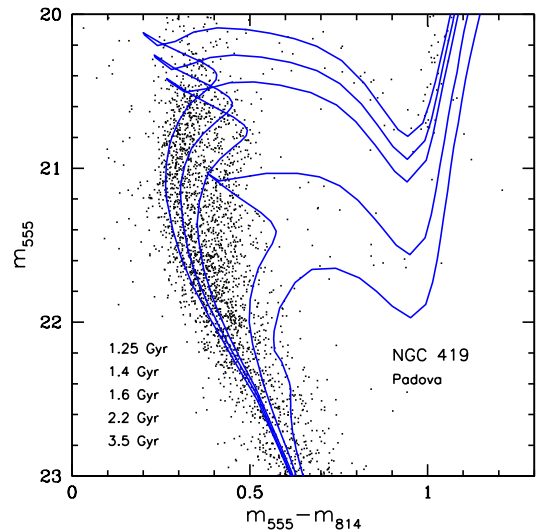}
 \caption{The CMD of NGC\,419 after zooming in to the region of the MSTO, SGB, and
 lower RGB. We show Padova isochrones as solid lines to estimate the age of the cluster and field population visible
 in the CMD. The isochrones cover an age range of 1.25, 1.4, 1.6, 2, and 3.5~Gyr. All other parameters are the same 
 as in Fig.~\ref{fig:n419_padova}.}
 \label{fig:n419_iso}
\end{figure}

In Figure~\ref{fig:n419_iso} we show Padova isochrones superimposed on the NGC\,419 region. These are isochrones 
of 1.25, 1.4, and 1.6~Gyr. Because the exact location of the MSTO of the cluster is uncertain, we do not give a 
single age for NGC\,419. Instead, we determine an age range of 1.2-1.6~Gyr. 

In Figure~\ref{fig:n419_teramo} we can clearly see the inability of the Teramo isochrones to fit all cluster 
features simultaneously. The best-fitting Teramo isochrone (1~Gyr) deviates increasingly to the blue along the 
MS. At the MSTO the isochrone is about 0.15~mag offset to the blue. The RGB turn-off is $\sim$0.2~mag too faint 
and the isochrone deviates again increasingly to the blue on the RGB.

\begin{figure}
  \epsscale{1}
  \plotone{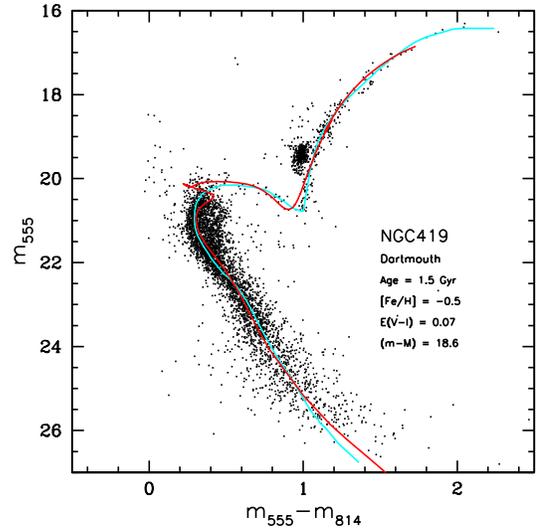}
 \caption{The CMD of NGC\,419 with the best-fitting Dartmouth isochrone overplotted in red. As before, the cyan line represents
 our fiducial for NGC\,419. The fit parameters are listed in the plot. }
 \label{fig:n419_dart}
\end{figure}

\begin{figure}
 \epsscale{1}
 \plotone{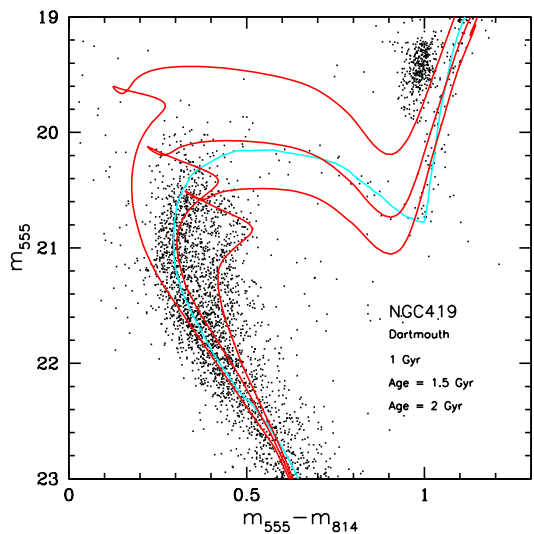}
 \caption{Same as Fig.~\ref{fig:n419_iso}, but for the Dartmouth isochrones with ages of 1, 1.5, and 2~Gyr.}
 \label{fig:n419_dartiso}
\end{figure}

The two youngest available Dartmouth isochrones have an age of 1 and 1.5~Gyr (Fig.~\ref{fig:n419_dart}). 
The age of NGC\,419 lies between 
these two isochrones. This might be the reason why the best-fitting Dartmouth isochrone does not provide a very 
good fit. The isochrone fits the RGB slope very well, but is $\sim$0.04~mag offset to the blue at the RGB 
turn-off. The isochrone is $\sim$0.25~mag too bright at the blue end of the SGB, and on the MS, the isochrone 
is slightly offset by about $\sim$0.02~mag to the red. In Figure~\ref{fig:n419_dartiso} we show the isochrones 
of 1 and 2~Gyr for comparison, which are obviously either too old or too young for NGC\,419. As for the Padova 
isochrones, we cannot give a single age for NGC\,419 and confirm the age range of 1.2-1.6~Gyr using the 
Dartmouth isochrones. Our derived reddening using the Dartmouth isochrone agrees with the extinction $A_V$ = 
$0.31$ from the \citet{Schlegel98} maps ($E_{V-I}$ = $0.08$~mag), while the reddening values found using the Padova 
and the Teramo isochrones are too high. 

Because of the complexity of NGC\,419, we will discuss its CMD in more detail in a separate paper. 

\section{Distances}
\label{sec:dist}

\begin{deluxetable*}{cccccc}
\tablecolumns{6}
\tablewidth{0pc}
\tablecaption{Distance Estimate}
\tablenote{The values for $V_{HB,RC}$ are taken from this paper and Paper~I. The values for $\langle M_V \rangle$ were
adopted from \citet{gir01}. For NGC\,121 and Lindsay\,38, we chose the model of Z=0.004, for Lindsay\,1, Kron\,3, NGC\,339, 
and NGC\,416 the model of Z=0.001, and for NGC\,419 the model of Z=0.004. The reddenings $E_{B-V}$ are taken from the Schlegel maps
\citep{Schlegel98} and transformed into $E_{V-I}$ by adopting $E_{V-I}/E_{B-V}$ = 1.25 from \citet{Dean78}. For NGC\,419 not
the luminosity of the red clump but the distance modulus found with the Padova isochrones ($(m-M)=18.75$) was used to determine
the distance due to the age and metallicity dependence of the absolute red clump magnitude.}
\tablehead{
\colhead{Cluster} & \colhead{$V_{HB,RC}$} & \colhead{$\langle M_V \rangle$} & \colhead{$E_{V-I,SM}$} & \colhead{$(m-M)$} &\colhead{Distance} \\
\colhead{} & \colhead{mag}  & \colhead{mag} & \colhead{mag} & \colhead{mag}  & \colhead{kpc}}
\startdata
NGC\,121   & $19.71 \pm 0.03$ & 0.574 & 0.024 & $19.06 \pm 0.03$ & $64.9 \pm 1.2$\\
Lindsay\,1 & $19.36 \pm 0.04$ & 0.509 & 0.024 & $18.78 \pm 0.04$ & $56.9 \pm 1.0$\\
Kron\,3    & $19.46 \pm 0.03$ & 0.474 & 0.024 & $18.91 \pm 0.04$ & $60.6 \pm 1.1$\\
NGC\,339   & $19.38 \pm 0.08$ & 0.455 & 0.040 & $18.80 \pm 0.08$ & $57.6 \pm 4.1$\\
NGC\,416   & $19.70 \pm 0.07$ & 0.474 & 0.104 & $18.90 \pm 0.07$ & $60.4 \pm 1.9$\\
Lindsay\,38& $19.60 \pm 0.05$ & 0.430 & 0.016 & $19.12 \pm 0.05$ & $66.7 \pm 1.6$\\
NGC\,419   & $19.41 \pm 0.12$ & -     & 0.080 & $18.50 \pm 0.12$ & $50.2 \pm 2.6$\\
\enddata
\label{tab:distance}
\end{deluxetable*}

There are no direct determinations of the cluster distances along the line-of-sight, but it is assumed that the SMC 
as a depth extent of up to 20~kpc \citep{Math88,Hatz93,crowl01,lah05}. We assume that the main body of the 
SMC has a distance modulus of $(m-M)_0$ = $18.88 \pm 0.1$~mag \citep[e.g., ][]{Storm04}. 

In addition to the distance estimates that result from our isochrone fits, we can use the apparent magnitudes of 
the red clump measured in this paper to provide a distance estimate for our clusters. We use the absolute red clump 
magnitudes $\langle M_V \rangle$ given in \citet{gir01}, as a function of age and metallicity. These authors 
provide mean properties of the red clump for metallicities from Z=0.0004 to 0.03, and ages from 0.5 to 12~Gyr, 
based on the theoretical horizontal branch models of \citet[][see also Girardi $\&$ Salaris 2001]{girar00}. We 
corrected our distance modulus for the interstellar extinction by consulting the Schlegel maps 
\citep{Schlegel98}. The adopted parameters and the resulting distances are listed in Table~\ref{tab:distance}. 
Due to the fact that NGC\,419 may be a multiple population object and therefore its age and the metallicity are 
uncertain, we do not use its absolute red clump magnitude for the distance estimate. Instead we apply the distance 
modulus found using the Padova isochrones (m-M) = 18.75~mag and correct for the extinction.

We find that Lindsay\,38 is the most distant cluster of our sample with a distance modulus $(m-M)$ = 19.12~mag 
($\sim$67~kpc), while NGC\,419 is the closest cluster with a distance modulus of $(m-M)$ = 18.50~mag ($\sim$50~kpc).  
NGC\,419 thus has a similar distance from us as the LMC, whose distance modulus is $18.50 \pm 0.02$~mag 
\citep[e.g., ][]{alves04}. Taking NGC\,419 into account, the closest and farthest cluster in our sample 
have a distance from each other of $\sim$17~kpc. Excluding NGC\,419, the depth extension of the SMC as derived from our 
cluster sample is $\sim$10~kpc. We have to emphasize that the distance of NGC\,419 is the most uncertain of our sample
due to its complexity.

\citet{crowl01} used the same approach and derived distances for 12 clusters, six of which
overlap with our sample. Their determined distance values are generally lower then the values we obtained, due to fainter 
absolute red clump magnitudes, which they adopted from \citet{sara99}. 
\citet{crowl01} do not have NGC\,419 in their sample, but NGC\,411, NGC\,152, Lindsay\,113, 
NGC\,361, Kron\,28, and Kron\,44. The closest cluster using the reddening values
from the Schlegel maps is Kron\,28 ($45.2 \pm 1.7$~kpc), and the farthest cluster is NGC\,121 ($65.4 \pm 1.9$~kpc).
Therefore Kron\,28 is $\sim$3~kpc closer than NGC\,419. In Crowl et al.'s sample, the clusters have a maximum distance 
of 20.2~kpc from one another, which is a higher value than what we have found with our smaller cluster sample. 
 
\section{Discussion}
\label{sec:ana}

\subsection{Comparison of our age determination with previous studies}
\label{sec:comp}

Previous studies done by several different authors provided ages and metallicities of SMC star clusters using a 
variety of techniques and telescopes. Therefore, if we combine all published cluster ages, we find a wide range 
of ages and metallicities for a given cluster, depending on the method used for the determination: Lindsay\,1 
has an age range from 7.3-10~Gyr \citep{gasc66,gasc80,ole87,sara95,migh98,udal98,alc03}, Kron\,3 from 5-10~Gyr 
\citep{gasc66,rich84,alc96,migh98,udal98,rich00}, and NGC\,339 from 5-7.9~Gyr \citep{migh98,udal98,rich00}. 

No other cluster has such a wide range of different age determinations as NGC\,416, 
reaching from 2.5-11.2~Gyr \citep{dur84, els85, bica86, migh98, udal98, rich00}. The cluster is located close to the
SMC main body where a large interstellar extinction is expected. The separation of field stars 
from the real cluster members was a major problem in the age determination process, among uncertain values for 
metallicity, reddening and distance. Using photometry obtained with the Wide Field Planetary Camera 2 (WFPC2) aboard
HST, \citet{migh98} found an absolute age of $6.6 \pm 0.5$~Gyr for NGC\,416, while 
\citet{rich00} derived an age of 7.1 to 11.2~Gyr using the same data set.

The only available CMD of Lindsay\,38 is provided by \citet{pia01}. The observation was carried out with the Cerro 
Tololo Inter-American Observatory (CTIO) 0.9~m telescope using the Tektronix 2K \#~3 CCD. They presented the first 
age determination of Lindsay\,38 with $6 \pm 0.6$~Gyr. For NGC\,419, the latest CMD was published by \citet{rich00} 
based on WFPC2 data. \citet{udal98} published an age of 3.3~Gyr and \citet{rich00} give an age range of $1.0-1.8$~Gyr.

For Lindsay\,1, Kron\,3, NGC\,339, NGC\,416, and NGC\,419, the latest and deepest available CMD was provided with WFPC2 
\citep{migh98,rich00}, while for Lindsay\,38 only ground-based data existed. 

\begin{deluxetable*}{ccccccc}
\tablecolumns{7}
\tablewidth{0pc}
\tablecaption{Age Comparison}
\tablenote{Comparison of our ages derived with the Dartmouth isochrones. For NGC\,419 Padova isochones provided the 
best fit. }
\tablehead{
\colhead{Ref.Source} & \colhead{Lindsay\,1} & \colhead{Kron\,3} & \colhead{NGC\,339} & \colhead{NGC\,416} & \colhead{Lindsay\,38} & \colhead{NGC\,419} \\
\colhead{} & \colhead{Gyr} & \colhead{Gyr} & \colhead{Gyr} & \colhead{Gyr} & \colhead{Gyr} & \colhead{Gyr}}
\startdata
This paper		   & $7.5 \pm 0.5$ & $6.5 \pm 0.5$ & $6 \pm 0.5$   & $6 \pm 0.5$   & $6.5 \pm 0.5$ & $1.2-1.6$     \\
\citet{rich00}  	   &	   -	   &   $5.6-7.9$   &  $5.0-7.9$    &   $4.0-7.1$   &	  -	   & $1.0-1.8$     \\
\citet{migh98}  	   & $7.7 \pm 0.4$ & $4.7 \pm 0.7$ & $5.0 \pm 0.6$ & $5.6 \pm 0.6$ &	-	   &	-	   \\ 
\citet{udal98}  	   &   $9.0$	   &  $7.5$	   &	$4.0$	   &  $6.6$	   &	  -	   &  $3.3$	   \\  
\citet{sara95}  	   & $7.3 \pm 0.6$ &	 -	   &	 -	   &	   -	   &	   -	   &	-	   \\ 
Alcaino et al. (1996, 2003) & $9-10$	   &  $8$	   &	 -	   &	   -	   &	   -	   &	-	   \\
\citet{pia01}		   &	   -	   &	  -	   &	   -	   &	    -	   & $6.0 \pm 0.6$ &	-	   \\
\enddata
\label{tab:age_com}
\end{deluxetable*}

In Table~\ref{tab:age_com} we compare our ages using the best-fitting Dartmouth isochrones (except NGC\,419, for 
which the Padova isochones provided the best fit) with results published in the most recent studies based on 
HST/WFPC2 photometry. The data reach $\sim$2~mag below the turnoff points, while our ACS data have a depth of 
3.5~mag below turnoffs. We can see that \citet{migh98} derived a similar age for Lindsay\,1, while for the 
remaining clusters in the overlapping sample they found younger ages than the ones derived here. \citet{rich00}, 
who used the same WFPC2 ''snapshots'' as \citet{migh98}, gave age ranges for certain metallicities for the clusters 
in their sample, which cover the ages determined in this paper. 

The ages published by \citet{udal98} using OGLE (Optical Gravitational Lensing Experiment) data, do not exhibit a 
general trend to older or younger ages as compared to our results, and the age difference varies for each cluster. 
The OGLE survey is a shallow ground-based survey with a limiting magnitude of $\sim$21~mag. \citet{sara95} used 
the B-V color difference between the HB and the RGB for star clusters with red HB morphologies for their age 
determination. The CMDs were obtained using data from the 2048 RCA prime-focus CCD on the CTIO 4~m telescope 
\citep{ole87} and the photometry reaches V$\sim$23~mag. The age found for Lindsay\,1 is in excellent agreement 
with our result. 

Alcaino et al. (1996, 2003) used photometry for Lindsay\,1 obtained with the 1.3~m Warsaw telescope, Las Campanas 
Observatory and reaches V$\sim$22~mag. The age was determined by using the so-called vertical method, based on 
the difference between the luminosity of the MSTO and the HB level. For Kron\,3, the photometry was taken 
with the EFOSC-2 CCD camera at the 2.2~m Max-Planck-Institute telescope of ESO , La Silla, and reaches V$\sim$23~mag 
\citep{alc96}. The age was determined using isochrones. In both studies, the resulting ages are higher than our 
values. \citet{pia01} were the first to publish an age for Lindsay\,38, which is in excellent agreement with the 
age derived in this paper.

Most CMDs published in previous studies do not go deep enough to show a clearly outlined MSTO, which is an 
essential feature for most age determination techniques. \citet{migh98} determined their cluster ages relative 
to the age of Lindsay\,1, measuring the difference between the RC and the RGB and found similar ages as in this 
paper. Kron\,3 is an exception for which the authors derived a younger age due to large error associated with the 
MS photometry. \citet{rich00} fitted isochrones to the red clump and also calculated the difference between the 
MSTO and the RC ($\Delta V^{RC}_{TO}$) in combination with the calibration of \citet{walker92}. The cluster ages 
found in this paper are within the age ranges given by \citet{rich00}. CMDs using ground-based photometry reach 
$\sim$V=20~mag, which is not deep enough to show the SGBs or the MSTOs, which can lead to large age differences.

\subsection{Age range and spatial distribution}
\label{sec:agerange}

The intermediate-age SMC star clusters Lindsay\,1, Kron\,3, NGC\,339, NGC\,416, and Lindsay\,38 form a continuous age 
sequence from 6 to 7.5~Gyr. The SMC is the only dwarf galaxy of the Local Group known to contain populous star clusters 
in this age range. The only ''true'' globular cluster in the SMC, NGC\,121, has an age of 10.5-11.5~Gyr (Paper~I), 
but is still 2-3~Gyr younger than the oldest LMC and MW globular clusters \citep[e.g., ][]{Ole96,olsen98,john99,mack04}. 
Between NGC\,121, and the second oldest cluster, Lindsay\,1, there is a small age-gap ($\sim$3~Gyr), in which no 
surviving star cluster has been formed. 

In our sample, we have four clusters with ages between 6-6.5~Gyr, and one that is significantly younger 
(1.2-1.6~Gyr). Good quality ages are available from ground-based and space-based observations for ten 
additional intermediate-age SMC star clusters. Combining them with our star clusters, we obtain a complete sample 
of all intermediate-age and old SMC star clusters: Kron\,28, Kron\,44, Lindsay\,116, Lindsay\,32, Lindsay\,11, 
NGC\,152, NGC\,361, NGC\,411, Lindsay\,113, and BS90 (Table~\ref{tab:lit_ages}).

\begin{deluxetable*}{ccccc}
\tablecolumns{5}
\tablewidth{0pc}
\tablecaption{Literature cluster ages}
\tablenote{Ages for nine additional intermediate-age clusters from the literature. }
\tablehead{
\colhead{Cluster} & \colhead{Age} & \colhead{Data} & \colhead{Method} & \colhead{Ref.Source}  \\
\colhead{} & \colhead{Gyr} & \colhead{} & \colhead{} & \colhead{}}
\startdata
Kron\,28     &  $2.1 \pm 0.5$ & CTIO 0.9~m telescope / Tektronix 2K \#~3 CCD& $\Delta V_{MSTO}^{RC,HB}$  & \citet{pia01}   \\
Kron\,44     &  $3.1 \pm 0.8$ & CTIO 0.9~m telescope / Tektronix 2K \#~3 CCD& $\Delta V_{MSTO}^{RC,HB}$  & \citet{pia01}   \\
Lindsay\,116 &  $2.8 \pm 1.0$ & CTIO 0.9~m telescope / Tektronix 2K \#~3 CCD& $\Delta V_{MSTO}^{RC,HB}$  & \citet{pia01}   \\
Lindsay\,32  &  $4.8 \pm 0.5$ & CTIO 0.9~m telescope / Tektronix 2K \#~3 CCD& $\Delta V_{MSTO}^{RC,HB}$  & \citet{pia01}   \\
Lindsay\,11  &  $3.5 \pm 1.0$ & CTIO 4.0~m telescope / RCA CCD  	    & Isochrones 		& \citet{mould92} \\
NGC\,152     &  $1.4 \pm 0.2$ & HST/WFPC2				    & Isochrones 		& \citet{crowl01} \\
NGC\,361     &  $8.1 \pm 1.2$ & HST/WFPC2				    & Isochrones 		& \citet{migh98}  \\
NGC\,411     &  $1.2 \pm 0.2$ & HST/WFPC2				    & Isochrones 		& \citet{alsa99}  \\
Lindsay\,113 &  $4.0 \pm 0.7$ & HST/WFPC2				    & $d_{B-V}$\tablenotemark{b} & Mighell et al. (1998a) \\
BS90	    &  $4.3 \pm 0.1$ & HST/ACS  				    & Isochrones 		& \citet{sabbi07} \\ 
\enddata
\tablenotetext{b}{The method used by Mighell et al. 
(1998a) is defined by the (B-V) color difference between the mean color of the red clump and the RGB at the level of
the RGB. This value then was compared with Lindsay\,1, NGC\,416, and Lindsay\,113.}
\label{tab:lit_ages}
\end{deluxetable*}

For none of these clusters deep HST photometry is available, thus their ages should be considered with 
some caution. For NGC\,152, NGC\,361, NGC\,411, Lindsay\,113, and BS90 ''snapshots'' are available taken with WFPC2 
(reaching V$\sim$23~mag), and ACS (BS90, reaching V$\sim$26~mag).

Looking at Figure~\ref{fig:clusters}, we clearly see that the youngest clusters are located near the SMC main 
body, while the clusters with ages higher than $\sim$4~Gyr lie in the outer parts. NGC\,361 seems to be an 
exception, but the cluster age is still uncertain, and the literature age of 8.1~Gyr probably is too high.  
\citet{crowl01} determined a distance of $51.7 \pm 1.8$~kpc for N361 whereby the cluster lies $\sim$ 7.5~kpc 
ahead of the SMC center. Another exception 
is BS90 that lies near the SMC main body, even though the cluster has an age of $\sim$4.3~Gyr. The three oldest 
SMC clusters (NGC\,121, Lindsay\,1, Kron\,3) are located in the north-western part of the SMC. We note that 
Lindsay\,116 cannot be seen in Figure~\ref{fig:clusters}, because it is located $6\arcdeg .1$ south-east of the 
bar and lies therefore outside the displayed area.


The closest cluster in our sample, NGC\,419, and the farthest cluster, Lindsay\,38, have a relative radial 
distance of 17~kpc from each other. We can therefore confirm that the SMC has a large extension along the 
line-of-sight, as was also found by \citet{crowl01} based on its star clusters.

\begin{figure}
 \epsscale{1.2}
 \plotone{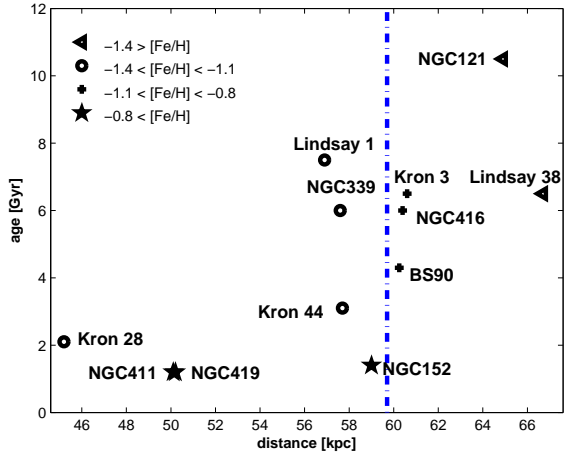}
 \caption{Age vs distance to the sun (projected distance) including different symbols for different metallicity ranges. For five clusters
 we found reliable distances \citet{crowl01}, ages \citep{alsa99,pia01,crowl01}, and metallicities (Kayser et al. 2008, 
 in prep.). All values for BS90 were adopted from \citet{sabbi07}. The dashed line represents  the SMC distance modulus of 
 $(m-M)_0$ = $18.88 \pm 0.1$~mag \citep{Storm04}.}
 \label{fig:distribution}
\end{figure}

\begin{figure}
 \epsscale{1.3}
 \plotone{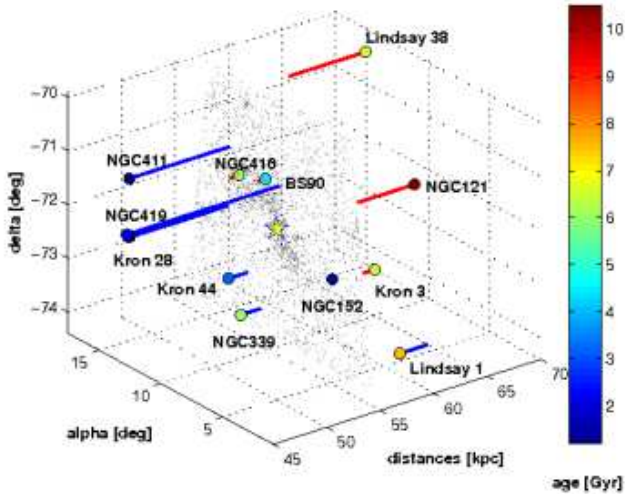}
 \caption{Three dimensional distribution is shown for SMC star clusters with ages and distances derived from isochrone 
 fits to CMDs derived from HST observations. Note that the intermediate age clusters are
 distributed throughout much of the extended body of the SMC. As discussed in the text, the selection of clusters is
 biased in that our observations generally avoided clusters in locations with high field star densities. However, this
 incomplete sample suggests that age and radial distance from the center of the SMC are not correlated; e.g. the 
 younger cluster Kron\,28, NGC\,411, and NGC\,419 are at large radial distances and cover a range in metallicity.
 The yellow star symbolizes the SMC center.}
 \label{fig:proj}
\end{figure}

In Figure~\ref{fig:distribution} we show the distribution of age vs the distance to the sun of the clusters in 
our sample. The locations are shown relative to our adopted SMC distance and indicate that the clusters
generally are distributed within $\pm$6-7~kpc of the SMC centroid. Interesting exceptions are the younger
clusters Kron\,28, NGC\,411, and NGC\,419 that in projection appear near the center of the SMC. In fact, they
could be located considerably closer to us (see also Fig.~\ref{fig:proj}). Further measurements of the distance
of younger clusters thus would be worthwhile. Moreover, we included five clusters for which we found reliable ages, 
distances, and metallicities in the literature. We divided the cluster metallicities into four groups and use 
different symbols for each group in the plot. Even though our plot contains only 11 clusters, we can see trends 
in the distributions of their properties. Age and distance from the sun appear to be correlated. The closest cluster, 
NGC\,419, is also the youngest and most metal rich cluster, while the most distant cluster, Lindsay\,38, is also 
the most metal poor, in spite of not being the oldest cluster. 

One could speculate that in regions at the outskirts of the double LMC-SMC system the star formation activity 
has been lower/slower than elsewhere, possibly with more unenriched gas, thus allowing for a more moderate 
enrichment. The oldest object, NGC\,121, is not the most metal poor cluster, but the second metal poorest and 
the second farthest one. Its low metallicity could be the result of both a ''natural'' age-metallicity relation 
and a ''distance from the system'' effect.

Figure~\ref{fig:proj} illustrates the distribution of SMC star clusters with high quality distances derived 
from isochrone fits to CMDs derived from HST observations. This is a highly biased sample; star clusters seen 
in the direction of the SMC `bar' are not preferred for these projects because of their large levels of field 
star contamination.  The exception in this case is the cluster BS90 that was accidently included in observations 
of NGC\,346 (see Sabbi et al. 2007). The present limited data for clusters in this project show that the SMC is 
quite extended along the line of sight, consistent with other studies of individual stars and star clusters 
(see discussion in $\S$~\ref{sec:dist}).  This three dimensional distribution of the clusters also demonstrates 
the lack of trends in cluster age or metallicity with radial distance from the center of the SMC.

\begin{deluxetable}{ccc}
\tablecolumns{3}
\tablewidth{0pc}
\tablecaption{Distances}
\tablenote{The projected distances were calculated in this paper and adopted from \citet{crowl01,sabbi07}.
Using these values and the cluster coordinates, we determined the cluster distances to the SMC center 
($\alpha = 0^h52^m44.8^s$, $\delta = -72\arcdeg49'43''$).}
\tablehead{
\colhead{Cluster} & \colhead{Projected Dist.} & \colhead{Dist. to SMC Center}  \\
\colhead{} & \colhead{kpc} & \colhead{kpc} }
\startdata
NGC\,121     & $64.9 \pm 1.2$ & $8.76 \pm 1.1$  \\
Lindsay\,1   & $56.9 \pm 1.0$ & $13.28 \pm 1.0$ \\
Kron\,3      & $60.6 \pm 1.1$ & $7.19 \pm 1.1$ \\
NGC\,339     & $57.6 \pm 4.1$ & $0.73 \pm 2.0$  \\
NGC\,416     & $60.4 \pm 1.9$ & $3.94 \pm 1.4$  \\
Lindsay\,38  & $66.7 \pm 1.6$ & $6.27 \pm 1.3$  \\
NGC\,419     & $50.2 \pm 2.6$ & $10.83 \pm 1.6$ \\
NGC\,411     & $50.1 \pm 1.7$ & $11.1 \pm 1.3$  \\
NGC\,152     & $59.0 \pm 1.8$ & $5.58 \pm 1.3$  \\
Kron\,28     & $45.2 \pm 1.7$ & $14.78 \pm 1.3$ \\
Kron\,44     & $57.7 \pm 1.8$ & $4.37 \pm 1.3$  \\
BS90	     & $60.3$	      & $1.23$  	\\
\enddata
\label{tab:lit_dist}
\end{deluxetable}

\subsection{Age distribution and cluster formation history}
\label{sec:agedistribution}

By combining the ages of our sample with 9 literature ages for intermediate-age SMC star clusters listed in  
Table~\ref{tab:lit_ages}, we obtain a well-observed sample of intermediate-age and old star clusters in 
the SMC. The cluster NGC\,361 was excluded from the sample, because the cluster is almost certainly younger 
than the assumed $\sim$8~Gyr \citep{migh98}.

The age distribution is shown in Figure~\ref{tab:age_hist}. In each panel we show our resulting age distribution 
using ages of different isochrone models (black histrograms) and the combined sample (white histograms). Since the 
cluster ages from the literature were derived using different data and methods, their distribution does not change. 

In all three plots of Figure~\ref{tab:age_hist}, the small age gap between $\sim$8 and 10~Gyr can clearly be seen. In 
the first panel we used ages derived with the Dartmouth isochrones. \citet{rich00} based on HST/WFPC2 found two brief 
cluster formation intervals with the oldest set $8 \pm 2$~Gyr ago and the second $2 \pm 0.5$~Gyr ago, and argued that 
there were gaps in between. During the older burst the clusters NGC\,339, NGC\,361, NGC\,416, and Kron\,3, and during 
the younger burst the clusters NGC\,411, NGC\,152, and NGC\,419 have formed according to \citet{rich00}. 

Even though they used the same HST/WFPC2 data as \citet{rich00}, \citet{migh98} found no evidence for such 
cluster formation bursts. We also find no evidence for two significant bursts of star cluster formation in our 
SMC age distribution, but we do see a slightly enhanced cluster formation activity around 6~Gyr. In the second 
and the third panel we used our derived Teramo and Padova ages, respectively. The cluster formation at 6~Gyr 
is even more obvious for both isochrone models than in the upper panel. 

Apparently, between $\sim$5 and 6~Gyr no star cluster with sufficient mass to survive has formed, but if Lindsay\,113 
is older than the assumed 4~Gyr adopted from the literature, the cluster lies within the gap. We suggest that 
the SMC has formed its clusters during its entire lifetime with some epochs of more intense cluster formation 
activity. More detailed information about the age distribution requires additional deep observations of all remaining 
intermediate-age SMC star clusters. 

As shown in Figure~\ref{fig:proj} there appears to be no simple relationship between cluster position and 
metallicity in any age range. This perhaps is to be expected given that tidal interactions may have perturbed 
the orbits of star clusters after they formed or provided opportunities for clusters to form at large radii, as 
in the present-day SMC wing. We have to emphasize that the cluster sample shown in Figure~\ref{fig:proj} is 
not complete. Only for 12 clusters reliable distances have been measured this far (see $\S$~\ref{sec:dist}), 
and these are shown in the Figure. The question of the metallicity distribution of the clusters and how this 
relates to age and position is more complex and beyond the scope of this paper. 

\subsection{Evolutionary history of the SMC as a whole}

Looking at the metallicities of our star clusters (Tab.~\ref{fig:results}), we see that the SMC did not 
experience a smooth age-metallicity relation, even though the SMC is believed to be well-mixed at
the present day \citep[but see ][]{Grebel92,gonzalez99}. The oldest SMC star cluster, NGC\,121, has a metallicity of 
[Fe/H] = $-1.46 \pm 0.10$ and an age of 10.5-11.5~Gyr, while Lindsay\,38 is more metal-poor with  [Fe/H] = 
$-1.59 \pm 0.10$ but has an age of $6.5 \pm 0.5$~Gyr. SMC star clusters of similar age may differ by several 
tenths of dex in metallicity (see also Da Costa et al. 1998, Kayser et al. 2008, in preparation). The probably 
most reasonable explanation involves the infall of unenriched, or less enriched gas. The Magellanic Clouds are 
surrounded by an extensive HI halo \citep[e.g.,][]{dickey96}, therefore this possibility may be plausible. 
Another speculative explanation for the existence of those metal-poor clusters is that the SMC acquired these 
clusters in a past interaction with another dwarf galaxy, similar to the clusters from the Sagittarius dwarf 
galaxy being acquired by the Milky Way \citep[e.g., ][]{carraro07}. 

The SMC, LMC, and MW form an interacting triple system, which affects each other's star formation history (SFH).
However, recent studies have suggested that the Magellanic Clouds only entered the vicinity of the MW fairly
recently (e.g., Kallivayalil et al. 2006a/b). It is intriguing that the LMC has a significant age gap between 
$\sim$4-9~Gyr, while the SMC formed its clusters continuously during the same time period. Moreover, the SMC 
appears to have a ''delayed'' globular cluster formation history and formed its first and only globular cluster, 
NGC\,121, 2-3~Gyr later then the LMC or the MW. 

\begin{figure}
  \epsscale{1.2}
  \plotone{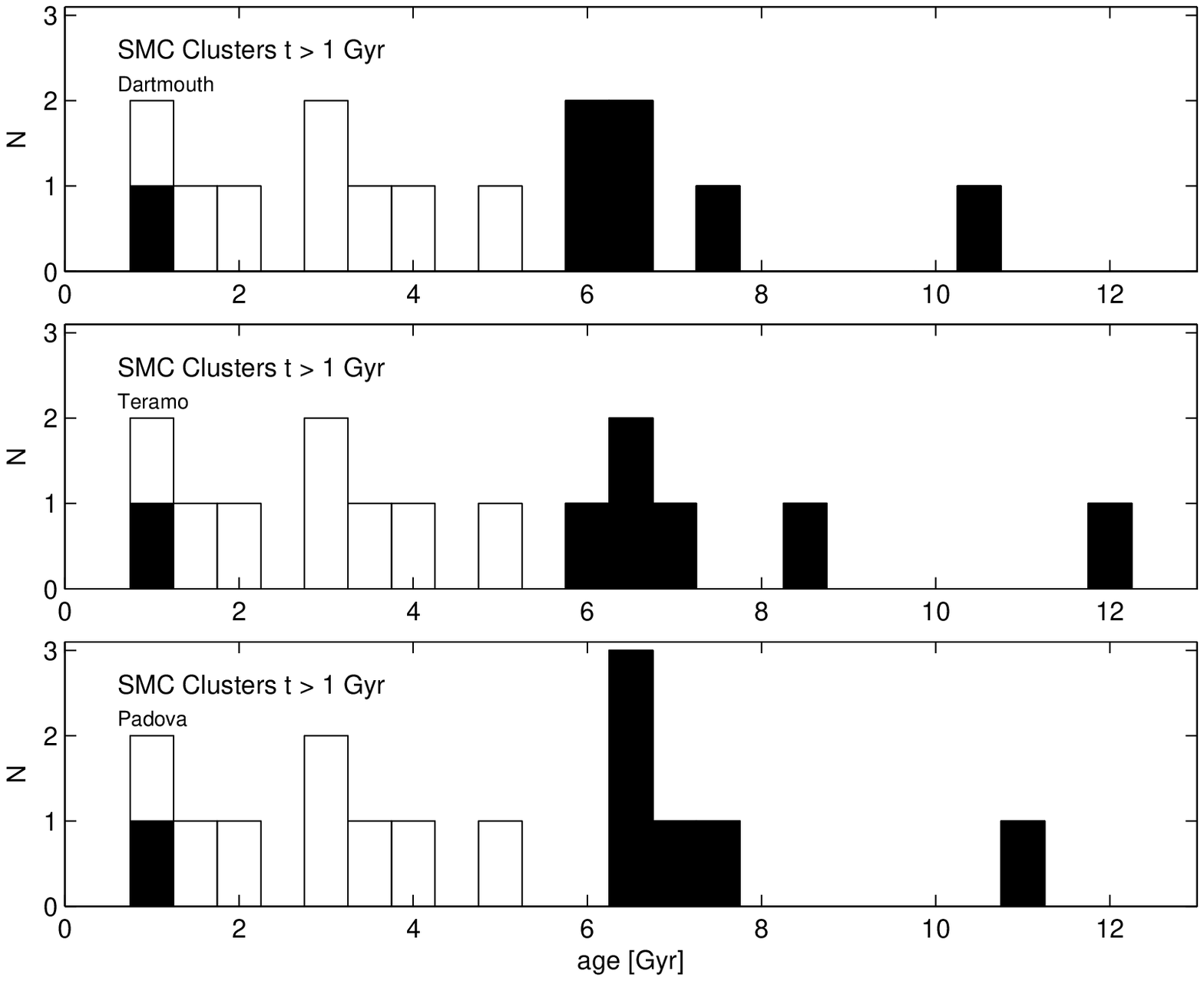}
 \caption{The age distribution of 15 intermediate-age and old SMC clusters (excluding NGC\,361) with ages derived in 
 this paper, in Paper~I (plotted as black histograms), and adopted from 
 \citep[Mighell et al. 1998a/b,][]{mould92,alsa99,rich00,pia01,crowl01} (plotted as white histrograms). Since the cluster
 ages from the literature were derived using different data and methods, their distribution does not change.
 In the first panel we used the 
 ages found using the Dartmouth models, in the second we use the Padova ages, and in the last panel the Teramo ages were 
 used. The literature ages of NGC\,152, NGC\,411, and Lindsay\,113, are based on HST/WFPC2 data 
 \citep[Mighell et al. 1998a/b,][]{alsa99,rich00}, 
 while the adopted ages of Kron\,28, Kron\,44, Lindsay\,11, Lindsay\,32, and Lindsay\,116 are derived from ground-based 
 photometry \citep{mould92,crowl01,pia01}. NGC\,361 is not considered due to its uncertain age. The age 
 distribution illustrates the continuous cluster formation with the small age gap between $\sim$8 and 10~Gyr.} 
 \label{tab:age_hist}
\end{figure}

Possible orbits of the SMC, LMC, and MW have been modelled by several authors 
\citep[e.g., Kallivayalil et al. 2006a/b;][]{Bekki05}. Strong tidal perturbations due to interactions could have 
triggered the cluster formation \citep[e.g., ][]{Whitmore99} in the SMC. In the LMC, we find that star clusters 
have formed in evident bursts. The LMC has two main epochs of cluster formation \citep[e.g., ][]{Bertelli92} and 
a well-known age-gap of several billion years, in which no star clusters have formed. A few globular clusters are 
found with coeval ages like the Galactic globular clusters \citep[e.g., ][]{ols91,olsen98,john99}. We know only 
of one star cluster, ESO 121-SC03, that lies within the age-gap, which has an age of 8.3-9.8~Gyr \citep{mackey06}. 
A correlation between young star clusters in the LMC and putative close encounters with the SMC and MW have been 
found by e.g. \citet{gir95}, although the most recent proper motion measurements indicate that the Magellanic 
Clouds are currently on their first passage around the MW. 

For young SMC clusters a relation between close encounters and the cluster formation history is not as obvious as 
for LMC clusters probably due to a smaller number of clusters \citep{Chiosi06}. The age distribution in 
Figure~\ref{tab:age_hist} shows that a slightly enhanced number of star clusters with ages around 1~Gyr is located 
in the SMC, which might have been produced through a cloud-cloud collision after a pericenter passage $\sim$0.5~Gyr 
ago \citep{Bekki05}. But evidently massive star clusters older than 1~Gyr have formed continuously 
until $\sim$7.5~Gyr ago (Lindsay\,1). It is not yet understood why populous star clusters older than 4~Gyr have 
not formed and survived continuously in the LMC, while in the SMC they did. \citet{Bekki05} explained the different 
cluster formation histories of the Clouds as a difference in birth locations and initial mass of the host galaxies. 

Kallivayalil et al. (2006a/b, see also Piatek et al. 2008) measured proper motions for the SMC and the LMC and used 
Monte Carlo simulations to model the orbits of the Clouds and the MW. While they found bound orbits for the Clouds, 
they also found that it was difficult to keep the Clouds bound to each other for more than 1~Gyr in the past. It 
is possible that the Clouds are not a bound system \citep[see also e.g., ][]{Bekki05}, and that they are making 
their first passage close to the Milky Way.

\section{SUMMARY}
\label{sec:sum}

\begin{deluxetable*}{ccccc}
\tablecolumns{5}
\tablewidth{0pc}
\tablecaption{Parameters}
\tablenote{All derived ages are listed.
The metallicities for the clusters NGC\,121, Lindsay\,1, Kron\,3 and NGC\,339 are taken from \citet{daco98}, where we adopted
the ZW84 metallicity scale. The metallicities for NGC\,416, Lindsay\,38, and NGC\,419, were taken from Kayser et al. 2008, 
in prep. in the CG97 scale, and which we transformed to ZW84 scale by using the transformation by \citet{cargra97}.}
\tablehead{
\colhead{Cluster} & \colhead{$[Fe/H]_{ZW84}$} & \colhead{$Age_{Teramo}$} & \colhead{$Age_{Padova}$} & \colhead{$Age_{Dartmouth}$} \\
\colhead{} & \colhead{} & \colhead{Gyr} & \colhead{Gyr} & \colhead{Gyr}}
\startdata
NGC\,121   & $-1.46 \pm 0.10$  & $11.8 \pm 0.7$ & $11.2 \pm 0.7$ & $10.5 \pm 0.5$ \\
Lindsay\,1 & $-1.14 \pm 0.10$  & $8.3 \pm 0.7$  & $7.7 \pm 0.7$ & $7.5 \pm 0.5$   \\
Kron\,3    & $-1.08 \pm 0.12$  & $7.2 \pm 0.5$  & $7.1 \pm 0.7$ & $6.5 \pm 0.5$ \\
NGC\,339   & $-1.12 \pm 0.10$  & $6.6 \pm 0.5$  & $6.3 \pm 0.5$ & $6 \pm 0.5$	\\
NGC\,416   & $-1.00 \pm 0.13$  & $6 \pm 0.8$	& $6.6 \pm 0.8$ & $6 \pm 0.8$	\\
Lindsay\,38& $-1.59 \pm 0.10$  & $6.3 \pm 0.5$  & $6.3 \pm 0.5$ & $6.5 \pm 0.5$\\
NGC\,419   & $-0.67 \pm 0.12$  & &1.2-1.6 & 1.2-1.6 \\
\enddata
\label{fig:results}
\end{deluxetable*}

In this paper, we have presented ages for the six intermediate-age SMC star clusters Lindsay\,1, Kron\,3, NGC\,339,
NGC\,416, Lindsay\,38, and NGC\,419 based on HST/ACS stellar photometry in the F555W and F814W passbands. The 
resulting CMDs represent the deepest published photometry so far and extend at least three magnitudes below the 
respective MSTOs. In order to obtain absolute ages, we 
applied three different isochrone models. The resulting ages are summarized in Table~\ref{fig:results}. We list the 
clusters by their identification in column (1). The [Fe/H] values are given in column (2) in the scale by 
\citet{zinn84}. Column (3) shows the magnitude of the MSTO $m_{555,TO}$, column (4) the magnitude of the red bump 
and column (5) the magnitude of the red clump $m_{555,RC}$. The columns (6), (7) and (8) show the absolute ages 
determined using the isochrone models of Teramo, Padova and Dartmouth.

We find that the Dartmouth isochrones provide the closest approximation to the MS, SGB, and RGB, whereas the other 
models mostly cannot reproduce the slope of the upper RGB when using the spectroscopically determined metallicity 
and requiring that the isochrones fit the MSTO and SGB. The Dartmouth isochrone models yield ages of $7.5 \pm 0.5$ 
for Lindsay\,1, $6.5 \pm 0.5$ for Kron\,3, $6 \pm 0.5$ for NGC\,339, $6 \pm 0.8$ for NGC\,416, and $6.5 \pm 0.5$ 
for Lindsay\,38. In general the isochrones provide good fits to the MSTO and SGBs that determine ages.

For the youngest cluster, NGC\,419, only the Padova isochrones fitted the CMD, while the Teramo isochrones had 
major problems with fitting the SGB and RGB, and the Dartmouth isochrones are not available for such young ages.

The difficulties of various isochrone models of given metallicities in reproducing the upper red giant branches 
of clusters with the same metallicities are a well-known problem \citep[e.g., ][]{greb97, greb99}. 
Figures~\ref{fig:l1_finaliso},~\ref{fig:k3_finaliso}, and~\ref{fig:ngc416_finaliso} reflect the general failure 
of the chosen stellar evolutionary models to simultaneously reproduce all of the major features of CMDs 
\citep[e.g., ][]{gall05} in spite of the excellent fit to the lower RGB, SGB, and MS. For each cluster we fitted 
a fiducial ridgeline that provides a unique set of low-metallicity fiducial isochrones, which are invaluable for 
detailed comparisons with stellar evolution models.

In each of our cluster CMDs stars blueward of and above the MSTOs are visible, which could be BSS. The radial
cumulative distribution of BSS candidats in Lindsay\,1, Kron\,3, NGC\,339, Lindsay\,38, and NGC\,419 showed that 
the stars found in the BSS regions are not concentrated toward the cluster centers and are therefore most probably 
part of the younger MS of the SMC field star population. For NGC\,416, we find an indication for centrally 
concentrated BSS candidates.

Looking at the spatial distribution, we find that the three oldest SMC clusters (NGC\,121, Lindsay\,1, Kron\,3) lie
in the north-western part of the SMC, while the youngest clusters are located near the SMC main body. Star
clusters with ages higher than $\sim$4~Gyr are located in the outer parts. 

From the observed red clump magnitude we give a distance estimate for our clusters. We find that Lindsay\,38 is 
the most distant cluster in our sample with d$\sim$68~kpc, while NGC\,419 is the closest cluster with  
d$\sim$53~kpc. Therefore, the closest and farthest cluster in our sample have a distance from each other of 
$\sim$17~kpc, which agrees with the assumed large depth extent of the SMC.

Further, we conclude that the SMC experienced massive cluster formation, remnants of which have survived
from over much of its lifetime, unlike the LMC or MW. The oldest and only globular cluster, NGC\,121, has 
formed $\sim$11~Gyr ago, while the next oldest set of surviving massive clusters, e.g. Lindsay\,1 or NGC\,361, 
date from approximately 3~Gyr later. After this time the largest age gaps are $\sim$1~Gyr suggesting that massive
star clusters occured without any substantial multi-Gyr hiatus. The SMC apparently formed massive star clusters 
that have survived from most of its lifetime.  
 
\acknowledgments
We would like to thank an anonymous referee for his or her useful comments. We gratefully 
acknowledge support by the Swiss National Science Foundation through grant 
number 200020-105260 and 200020-113697. Support for the US component of this program GO-10396 
was provided 
by NASA through a grant from the Space Telescope Science Institute, which is operated 
by the Association of Universities for Research in Astronomy, Inc., under NASA contract 
NAS 5-26555. We warmly thank Paolo Montegriffo to
provide his software and Leo Girardi for the Padova isochrones in the standard ACS color 
system. Gisella Clementini and Monica Tosi have been partially supported by PRIN-MIUR-2004 
and PRIN-INAF-2005, and Jay Gallagher also obtained helpful additional support from the 
University of Wisconsin Graduate School and from the Heidelberg Graduate School of
Fundamental Physics.


\end{document}